\documentclass[a4paper,aps,twocolumn,nofootinbib]{revtex4}
\RequirePackage[colorlinks,hyperindex]{hyperref}
\RequirePackage[english]{babel}
\RequirePackage[latin1]{inputenc}
\RequirePackage[T1]{fontenc}
\RequirePackage{mathrsfs}
\RequirePackage{amsfonts}
\RequirePackage{amsmath}
\RequirePackage{amssymb}
\RequirePackage{amsbsy}
\RequirePackage{color}
\RequirePackage{bm}
\usepackage{tensind}
\usepackage[a4paper, total={7in, 10in}]{geometry}
\newcommand{\comm}[1]{}
\newcommand{\Tr}{{\rm Tr}\,}
\newcommand{\Det}{{\rm Det}\,}
\DeclareMathOperator{\arctanh}{arctanh}
\DeclareMathOperator{\csch}{csch}
\hypersetup{colorlinks=true,breaklinks=true,urlcolor=blue,linkcolor=red}
\begin{document}
\title{\bf{Relativistic dynamical inversion in manifestly covariant form}}
\author{Andre G. Campos}
\email{agontijo@mpi-hd.mpg.de}
\affiliation{Max Planck Institute for Nuclear Physics, Heidelberg 69117, Germany} 

\author{Luca Fabbri}
\email{fabbri@dime.unige.it}
\affiliation{DIME, Sez. Metodi e Modelli Matematici, Universit\`{a} di Genova, 
Via all'Opera Pia 15, 16145 Genova, ITALY} 
\date{\today}
\begin{abstract}
The Relativistic Dynamical Inversion technique, a novel tool for finding analytical solutions to the Dirac equation, is written in explicitly covariant form. It is then shown how the technique can be used to make a change from cartesian to spherical coordinates of a given Dirac spinor. Moreover the most remarkable feature of the new method, which is the ease of performing non-trivial change of reference frames, is demonstrated. Such a feature constitutes a potentially powerful tool for finding novel solutions to the Dirac equation. Furthermore, a whole family of normalizable analytic solutions to the Dirac equation is constructed. More specifically, we find exact solutions for the case of a Dirac electron in the presence of a magnetic field as well as a novel solution comprising of a combination of a spherically symmetric electric field and magnetic fields. These solutions shed light on the possibility of separating the positive and negative energy parts of localized Dirac spinors in the presence as well as in the absence of magnetic fields. The presented solutions provide an illustration of the connection between the geometrical properties of the spinor and spin-orbit coupling for normalizable spinorial wave functions.

\end{abstract}
\maketitle
\section{Introduction}
The \textit{Relativistic Dynamical Inversion} technique (or RDI for short) first proposed in \cite{RDI1}, was designed to solve the following problem: Given an arbitrary (desired) spinorial spacetime wave packet $\psi$, find an electromagnetic vector potential $A_{\mu}$ such that the Dirac equation is satisfied. Several examples demonstrating the ability of RDI to find novel non-trivial analytical solutions to the Dirac were discussed in \cite{RDI1,RDI2,RDI3} . The most appealing feature of RDI is the clear geometrical meaning bestowed upon the spinor by the Hestenes formalism on which it is based, whose main advantage is that a clear classical interpretation can be given to the elements of the theory; this in turn provides the intuition of how the electron in the quasi-classical approximation is expected to move in the desired field configuration for which the solutions to the Dirac equation are sought, with the information about such ``motion'' being encapsulated in the geometrical interpretation of $\psi$. However, in its current form, RDI is formulated in cartesian coordinates, which severely hinders the feasibility of the technique to tackle problems possessing certain symmetries, for instance spherical or cylindrical symmetry, which generally lead to a great deal of simplification in the equations one need to solve. A general way to address this issue is to rewrite the equations of RDI in manifestly covariant form. Another advantage in making RDI explicitly covariant is that only then can gravity also be included. 

Before proceeding, it is noteworthy that Hestenes (see \cite{hestenes2020spacetime} and references therein) already put forward a covariant version of his technique, which should be equivalent to the one developed here. However, while Hestenes' formulation relies heavily on his so called geometric algebra, with a plethora of new symbols and calculation rules, the formalism developed here relies only on basic matrix algebra and tensor calculus, consequently being accessible to a wider audience. Moreover, Hestenes focused mainly on gravitational effects which might mislead the reader to thinking that it is all his formalism is about; in contrast, our main focus is on the covariance of the formulation, which applies not only to curved spacetime, but also to curvilinear coordinates and non-inertial frames in flat spacetime. In addition, our technique can also deal with spacetimes endowed with torsion.

In this work we put forward the \textit{Manifestly Covariant Relativistic Dynamical Inversion} (CRDI) technique, a novel formalism which is the offspring of the marriage between the Hestenes formulation with the formalism of \textit{Polar Spinors} (see, for instance, \cite{Fabbri:2020ypd,Fabbri:2021mfc}).
In order to illustrate the usefulness of CRDI, we start with a general form of the spinorial spacetime wave packet $\psi$ constituting a general solution to the Dirac equation in cartesian coordinates and show that a particular case of such solution is the ground state of the Hydrogen atom. Then we make a change from cartesian to spherical coordinates and demonstrate a novel feature of CRDI, which is the easiness of performing non-trivial change of reference frames that can potentially be a powerful tool in the quest of novel analytical solutions to the Dirac equation. In addition, we construct a whole family of normalizable analytic solutions to the Dirac equation. More specifically, we find exact solutions for the case of a Dirac electron in the presence of a magnetic field as well as a novel solution consisting of the combination of a spherically symmetric electric field and magnetic fields. These solutions give some clues on the relationship between magnetic fields, spin-orbit coupling and the geometrical properties of the Dirac spinor.

This paper is organized as follows. In Sec. \ref{GenDesc} an overview of the steps needed to write the Dirac equation in explicitly covariant form is given. In Sec. \ref{RDI} we summarize both the Hestenes formalism and the RDI technique. Then we present a step by step derivation of CRDI in Sec. \ref{CRDI}. In Sec. \ref{illustration} we first show how the chosen general $\psi$ reproduces the solutions presented in \cite{RDI3} and then proceed to describe two families of analytical solutions of the Dirac equation, one of which is novel, constructed with the help of our newly developed technique. In Sec. \ref{discussion} we analyze the newly found solutions in light of the geometrical interpretation of CRDI. Finally, in Sec. \ref{conclusion} we draw some conclusions and highlight some of the potential applications of CRDI.

\section{General Covariance and Dirac Spinors}\label{GenDesc}
In this section we describe the steps needed to write the Dirac equation in manifestly covariant form. This discussion is based on the procedure introduced in \cite{utiyama1956invariant}.

Hereafter, Greek indices label the spacetime manifold while Latin indices label the tangent space to the manifold. Since in Sections \ref{HestenesF} and \ref{Baylis} there is no distinction between Greek and Latin indices (everything is written in cartesian coordinates), we use Greek indices throughout both sections. 

The general procedure to write the Dirac equation in explicitly covariant form is as follows:
choose a single point on a curved manifold. There the manifold is exactly equal to the tangent space at the chosen point. Since we are considering the Lorentzian curved metric tensor, the tangent space at every point will be Minkowski. One can then define a field at each point that transforms as a scalar under coordinate transformations in the curved manifold, and as a spinor under the Lorentz group in the tangent space. Given that at the chosen point the curved manifold and the tangent space are identical, one has just constructed a Lorentz spinor at one point on a curved manifold. One may then construct such an object at every other point.

Each point on the curved manifold has its own independent tangent space in which it can transform under an arbitrary Lorentz transformation. Thus, to simultaneously define a Lorentz spinor at all points on a curved manifold, one must let it transform under local Lorentz transformation in each independent tangent space at each point. Therefore the first step to write the Dirac equation in covariant form is to construct the complexification of the $\mathrm{SO(1,3)}$ gauge theory for spinors. 

To identify the space of $\mathrm{SO(1,3)}$ gauge transformations with the space tangent to the manifold at each point, one must do the following: First, as stated at the beginning of this section, one must identify the $\mathrm{SO(1,3)}$ indices (Latin indices) with the tangent space Lorentz indices. Second, one must find a way to define an independent set of gamma matrices at each point on the curved manifold which act at the tangent space at that point. And third, one must find a way to relate the spin connection to the geometry of the curved manifold such that a tangent space Lorentz transformation under which the theory must be invariant simply translates to a $\mathrm{SO(1,3)}$ gauge transformation on the spinor field.
 
The first task can be straightforwardly taken care of by simply identifying the Latin indices with flat Lorentz indices. The other two are more involved. It turns out that the first step towards simultaneously tackling both of them is to introduce a new field that relates the curved metric with the tangent space metric. It is called the tetrads, commonly denoted in the literature by the symbols $e^\mu_a$ and $e^a_\mu$, obeying the following relations 
\begin{align}
\eta_{ab}e^a_\mu e^b_\nu = g_{\mu\nu}
\end{align}
where $\eta_{ab}$ is the Minkowski metric on the flat tangent space while $g_{\mu\nu}$ is the curved metric tensor on the manifold. The geometrical meaning of the tetrads is that they constitute a set of four orthonormal vectors (in terms of the Minkowski metric) at each point on the tangent space to the manifold. The definition of the gamma matrices as matrix valued functions of the coordinates on the manifold is done by imposing that they must obey the following curved Clifford algebra
\begin{equation}
\{\gamma^\mu,\gamma^\nu\}=2g^{\mu\nu}\boldsymbol{1}
\label{1}
\end{equation}
where $\boldsymbol{1}$ is the $4\times4$ identity matrix and $\{.,.\}$ is the anticommutator. The tetrad is relevant here because the solution to the above anticommutation relation can simply be written in terms of the standard flat gamma matrices using the tetrad, that is 
\begin{equation}
e^a_\nu\gamma_a=\gamma_\nu. 
\end{equation}
Finally the third and final task is to write the spin connection in terms of the geometry of the curved manifold so that tangent space Lorentz transformations under which the theory must be invariant simply translate to $\mathrm{SO(1,3)}$ gauge transformation, under which the theory is already invariant. This can be done, it turns out, by using the so called metronilic property, $\boldsymbol{\nabla}_\mu\gamma_\nu=0$. This task is more involved and will be addressed in Sec. \ref{HestenesCovIntro}.
\section{An introduction to RDI}\label{RDI}
In this section we give a brief review of both the Hestenes formalism and the RDI technique.

\subsection{A review of the Hestenes formalism}\label{HestenesF}
This section is a reinterpretation of the work of Hestenes \cite{hestenes1975observables} in terms of the quaternion formalism developed in Ref. \cite{gursey1956contribution}. The new point of view consists in interpreting $\gamma_\mu$ as vectors of a space-time reference frame instead of matrices. By definition the scalar product of these vectors is just the components $\eta_{\mu\nu}$ of the metric tensor
\begin{align}\label{dotProd}
\frac{1}{2}(\gamma_\mu\gamma_\nu+\gamma_\nu\gamma_\mu)=\gamma_\mu\cdot\gamma_\nu=\eta_{\mu\nu}
\end{align}
generating an associative algebra over the real numbers, which has been called the \textit{space-time algebra} by Hestenes. It provides a direct and complete algebraic characterisation of the geometric properties of Minkowski space-time in cartesian coordinates.
In the standard representation the Dirac equation for an electron with charge $e$ and mass $m$ in an external electromagnetic field $A_\mu$ reads
\begin{align}\label{standardDirac}
\gamma^\mu\left(i\hbar\partial_\mu-eA_\mu\right)\psi=mc\psi.
\end{align}
The Dirac spinor $\psi\in\mathbb{C}^4$ obeying (\ref{standardDirac}) is a column vector with four complex components
\begin{align}\label{diracColum}
\psi=\begin{pmatrix}
 \psi_1 \\ \psi_2 \\ \psi_3 \\ \psi_4
 \end{pmatrix}=\begin{pmatrix}
 r_0-ir_3 \\ r_2-ir_1 \\ \mathfrak{s}_3+i\mathfrak{s}_0 
 \\ \mathfrak{s}_1+i\mathfrak{s}_2
 \end{pmatrix},
\end{align}
where the $\mathfrak{s}_\mu$ and $r_\mu$ are real functions of space-time. The representation (\ref{diracColum}) in terms of the components $\psi_1,\psi_2,\psi_3,\psi_4$ presumes a specific representation of the Dirac matrices, the standard (Dirac) representation, 
\begin{align}\label{diracmatrices}
\gamma_0=\begin{pmatrix}
 I & 0 \\
 0 & -I 
 \end{pmatrix},\quad\gamma_k=\begin{pmatrix}
 0 & -\sigma_k \\
 \sigma_k & 0 
 \end{pmatrix}
\end{align}
subject to $\gamma^{0}(\gamma^{\mu})^\dagger\gamma^{0}=\gamma^{\mu}$ as well as $\gamma^{2}(\gamma^{\mu})^{*}\gamma^{2}=\gamma^{\mu}$ where $I$ is the $2\times2$ identity and $\sigma_k$ the Pauli matrices.

The matrices  $\alpha_k=\gamma_k\gamma_0$ are to be interpreted as unit quaternions. The $\alpha_k$ generates an algebra over the real numbers which is isomorphic to the Pauli algebra. This fact is emphasized by writing
\begin{align}\label{STA}
\alpha_1\alpha_2\alpha_3=\gamma_0\gamma_1\gamma_2\gamma_3=\boldsymbol{i},\quad\boldsymbol{i}^2=-1.
\end{align}
Thus, $\boldsymbol{i}$ plays a similar role as $i=\sqrt{-1}$ does in the Pauli Algebra. In quaternion theory, it is an additional operator which commutes with the $\alpha_k$ and squares to $-1$.
From the standard representation (\ref{diracmatrices}) it follows
\begin{align}
\nonumber
\alpha_k=\begin{pmatrix}
 0 & \sigma_k \\
 \sigma_k & 0 
 \end{pmatrix},\quad\boldsymbol{i}=\begin{pmatrix}
 0 & iI \\
 iI & 0 
 \end{pmatrix},\\ \boldsymbol{i}\alpha_k=\begin{pmatrix}
 i\sigma_k & 0\\
 0& i\sigma_k 
 \end{pmatrix},\quad \gamma^5=i\gamma^0\gamma^1\gamma^2\gamma^3=\begin{pmatrix}
 0 & I \\
 I & 0 
 \end{pmatrix}\label{gamma5}.
\end{align}

By introducing the canonical basis in the spinor space
\begin{align}\label{spinorBasis}
u_1&=\begin{pmatrix}
 1 \\ 0 \\ 0 \\ 0
 \end{pmatrix},\ u_2=\begin{pmatrix}
 0 \\ 1 \\ 0 \\ 0
 \end{pmatrix},\ u_3=\begin{pmatrix}
 0 \\ 0 \\ 1 \\ 0
 \end{pmatrix},\ u_4=\begin{pmatrix}
 0 \\ 0 \\ 0 \\ 1
 \end{pmatrix}, \\
\gamma_0u_1&=u_1, \,\boldsymbol{i}\alpha_3 u_1=\gamma_2\gamma_1u_1=iu_1,\label{a}\\
 u_2&=-\boldsymbol{i}\alpha_2u_1,\, u_3=\alpha_3u_1,\quad u_4=\alpha_1u_1\label{c},
\end{align}
we have that the Dirac spinor $\psi$ given in (\ref{diracColum}) in this representation can be written as
\begin{eqnarray}
\nonumber
&\psi=\psi_1u_1+\psi_2u_2+\psi_3u_3+\psi_4u_4\\
\nonumber
&=[r_0\boldsymbol{1}+\left(\mathfrak{s}_1\alpha_1+\mathfrak{s}_2\alpha_2+\mathfrak{s}_3\alpha_3\right)-\\
&-\boldsymbol{i}\left(r_1\alpha_1+r_2\alpha_2+r_3\alpha_3\right)+\boldsymbol{i}\mathfrak{s}_0]u_1
\label{hestenesS}
\end{eqnarray}
where relations (\ref{a}) and (\ref{c}) have been used.
Thus any Dirac spinor $\psi$ can be written as
\begin{align}
\label{DiracS2}
\psi=\Psi u_1
\end{align}
where $\Psi$ can be written down directly from the column matrix form (\ref{diracColum}) by using
\begin{eqnarray}
\nonumber
&\Psi=r_0\boldsymbol{1}+\left(\mathfrak{s}_1\alpha_1+\mathfrak{s}_2\alpha_2+\mathfrak{s}_3\alpha_3\right)-\nonumber\\
&-\boldsymbol{i}\left(r_1\alpha_1+r_2\alpha_2+r_3\alpha_3\right)+\boldsymbol{i}\mathfrak{s}_0=\nonumber\\
&=\begin{pmatrix}
 \psi_{1} & - \psi_2^{*} & \psi_3 & \psi_4^{*} \\
 \psi_2 & \psi_1^{*} & \psi_4 & -\psi_3^* \\
 \psi_3 & \psi_4^* & \psi_1 & -\psi_2^* \\
 \psi_4 & -\psi_3^* & \psi_2 & \psi_1^* 
 \end{pmatrix}
 \label{HestenesG}.
\end{eqnarray}
The matrix spinor $\Psi$ is then a general complex quaternion.

The most general form of the spinor $\Psi$ can be translated into the polar form
\begin{align}\label{GeneralMatrixPsi}
 \Psi = \sqrt{\rho} \exp \left(\boldsymbol{i} \beta/2 \right)
\mathcal{R}, 
\end{align}
where $\mathcal{R}=\mathcal{B}U$ \cite{hestenes1967real}. The matrices $U$ and $\mathcal{B}$ are unitary and Hermitian, respectively, with $U$ encoding the rotations and $\mathcal{B}$ performing the boosts.
From equation (\ref{GeneralMatrixPsi}) we have, noting that $U\gamma_0=\gamma_0U$
\begin{align}
\label{velocity1}
\Psi\gamma_0\tilde{\Psi}=\rho v\!\!\!/=\rho \mathcal{B}\gamma_0\mathcal{B}^{-1}=\rho \mathcal{B}^2\gamma_0
\end{align}
since $\mathcal{B}^{-1}=\tilde{\mathcal{B}}=\gamma_0\mathcal{B}^\dagger\gamma_0=\gamma_0 \mathcal{B}\gamma_0$. Then
\begin{align}
\label{spin1}
\Psi\gamma_3\tilde{\Psi}=\rho s\!\!\!/=\rho \mathcal{B}U\gamma_3U^\dagger \mathcal{B}^{-1}=\rho \mathcal{B} U\alpha_3 U^\dagger \mathcal{B}\gamma_0
\end{align}
for the spin. 
Before proceeding, we would like to address the issue of why is the vector spin density in (\ref{spin1}) calculated with $\gamma_3$ and does
not involve $\gamma_5$. The reason is that it is a feature of the Hestenes formalism; in such a formalism the $\gamma_5$
is somewhat hidden in the definition of the matrix spinor while the $\gamma_3$
appears because the spin is aligned along the third axis. However,
these are only `incidents' of the choice of formalism. Of course when
written in the usual formalism the $\gamma_3$ leaves its place to the $\gamma_5\gamma_3$ and
all the usual definitions are recovered. Finally, we would like to call attention to the
following definitions
\begin{equation}
\Psi^{-1}=\frac{\tilde{\Psi}}{\Psi\tilde{\Psi}},\, \tilde{\Psi}=\gamma_0\Psi^\dagger\gamma_0,\, \Psi\tilde{\Psi}=\rho e^{\boldsymbol{i}\beta},\,\mathcal{R}^{-1}=\tilde{\mathcal{R}}.
\end{equation}

The general matrix spinor (\ref{GeneralMatrixPsi}) satisfies the Hestenes-Dirac equation
\begin{align}
 \left(\hbar c \partial\!\!\!/ \Psi \gamma_2\gamma_1 - c q A\!\!\!/ \Psi\right) = m c^2 \Psi\gamma_0\label{Dirac-Hestenes-Eq4}
 \end{align}
where $A\!\!\!/ = A^{\mu} \gamma_{\mu}$ and $\partial\!\!\!/ = \gamma^{\mu} \partial_{\mu}$. The factorization (\ref{GeneralMatrixPsi}) implies
\begin{align}\label{FactorizationD}
\partial\!\!\!/\Psi=\frac{1}{2}\left(\partial\!\!\!/\ln{\rho}+\partial\!\!\!/\beta\boldsymbol{i}-2\gamma^\mu\mathcal{R}\partial_\mu\mathcal{R}^{-1}\right)\Psi.
\end{align}
Substituting (\ref{FactorizationD}) in (\ref{Dirac-Hestenes-Eq4}) leads to
\begin{eqnarray}
\nonumber
&\!\!\!\!\hbar(\partial\!\!\!/\ln{\sqrt{\rho}}\!+\!\frac{\partial\!\!\!/\beta\boldsymbol{i}}{2}-\gamma^\mu\mathcal{R}\partial_\mu\mathcal{R}^{-1})\Psi\gamma_2\gamma_1-\\
&-qA\!\!\!/\Psi\!=\!mc\Psi\gamma_0.
\label{Dirac-Hestenes-Eq5}
\end{eqnarray}
\subsection{Summary of the Relativistic dynamical inversion technique}
Once the matrix spinor is given, the next step is to find the electromagnetic fields that induce the motion of the electron encoded in $\Psi$. Formally, the vector potential can be written in terms of $ \Psi $ by inverting (\ref{Dirac-Hestenes-Eq5}) as
\begin{align}
 e A\!\!\!/= \hbar \partial\!\!\!/ \Psi  \gamma^2\gamma^1\Psi^{-1} -  m c \Psi \gamma^0\Psi^{-1},
   \label{GeneralA}
\end{align}
The vector potential equation can also be rewritten in a more illuminating form
\begin{align}
 e A\!\!\!/= \hbar \partial\!\!\!/ \Psi  \gamma^2\gamma^1\Psi^{-1} -  p\!\!\!/e^{-\boldsymbol{i}\beta},\quad  p\!\!\!/=mc v\!\!\!/,
   \label{GeneralA2}
\end{align}
which allows us to identify $p\!\!\!/$ with the  kinetic momentum.

The vector potential given by (\ref{GeneralA}) is required to obey the following constraints
\begin{align*}
&\Tr[e A\!\!\!/\Gamma_1]/4=0,\nonumber\\
&\Tr[e A\!\!\!/\Gamma_n]/4=0,\mbox{ for } 6\leq n\leq 16,
\end{align*}
where
\begin{align*}
&\Gamma_1=\boldsymbol{1},\quad \Gamma_2=\gamma^0,\quad\Gamma_3=\gamma^1,\quad\Gamma_4=\gamma^2,\\
&\Gamma_5=\gamma^3,\quad\Gamma_6=\alpha_1,\quad\Gamma_7=\alpha_2,\quad\Gamma_8=\alpha_3,\nonumber\\
&\Gamma_9=\gamma^2\gamma^3,\quad \Gamma_{10}=\gamma^3\gamma^1,\quad\Gamma_{11}=\gamma^1\gamma^2,\\
&\Gamma_{12}=\gamma^1\gamma^2\gamma^3,\quad\Gamma_{13}=\gamma^0\gamma^2\gamma^3,\quad\Gamma_{14}=\gamma^0\gamma^3\gamma^1,\nonumber\\
&\Gamma_{15}=\gamma^0\gamma^1\gamma^2,\quad\Gamma_{16}=\gamma^5.
\end{align*}
The above conditions imply
\begin{align}\label{constraints}
&\partial_\mu (\rho v^\mu)=0,\nonumber\\
&\partial_\mu(\rho s^\mu)+\frac{2mc}{\hbar}\rho\sin\beta=0.
\end{align}

Thus, after taking into account the constraints (\ref{constraints}), the components of the vector potential such that the Hestenes-Dirac equation is satisfied by the given matrix spinor $\Psi$ are 
\begin{align}
eA_0&=\frac{\hbar}{2}\Bigg(-\frac{1}{v_0^2}[\boldsymbol{s}+\boldsymbol{v}\times(\boldsymbol{s}\times\boldsymbol{v})]\cdot\vec{\nabla}\beta-\boldsymbol{e}_2\cdot\partial_0\boldsymbol{e}_1\nonumber\\
&+\frac{\vec{\nabla}\cdot(\rho\boldsymbol{s}\times\boldsymbol{v})}{\rho}\Bigg)-mcv_0\cos\beta\label{scalarpart},\\
eA_k&=\frac{\hbar}{2}\Bigg(-v_ks_\mu\partial^\mu\beta+s_kv_\mu\partial^\mu\beta-\boldsymbol{e}_2\cdot\partial_k\boldsymbol{e}_1\nonumber\\
&+\frac{1}{\rho}\Bigg[\varepsilon_{klm}\frac{\partial}{\partial x^l}(\rho\{\frac{1}{v_0^2}[\boldsymbol{s}+\boldsymbol{v}\times(\boldsymbol{s}\times\boldsymbol{v})]^m\})\nonumber\\
&-\varepsilon_{klm}\frac{\partial}{c\partial t}(\rho s^lv^m)\Bigg]\Bigg)-mcv_k\cos\beta\label{vectorpart}.
\end{align}
where $\boldsymbol{e}_{j}=\Psi\gamma_{j}\tilde{\Psi}/\rho$, $j=1,2$, $\boldsymbol{e}_2\cdot\partial_0\boldsymbol{e}_1=\Tr[\boldsymbol{e}_2\partial_0\boldsymbol{e}_1]/4$ and $\boldsymbol{e}_2\cdot\partial_k\boldsymbol{e}_1=\Tr[\boldsymbol{e}_2\partial_k\boldsymbol{e}_1]/4$.
\subsection{Form Hestenes to Baylis: anatomy of the matrix spinor}\label{Baylis}
The first step towards gaining more physical insight is the equivalence between the Baylis formulation known as \textit{Algebra of Physical Space} (APS) \cite{BaylisClassicalSpinor1992,BaylisBook1996} and the Hestenes formalism introduced in Sec. \ref{HestenesF}; to the best of our knowledge, this equivalence was previously discussed only by G\"ursey in \cite{gursey1956contribution}.

In the Baylis formalism the state $\psi$ can be represented by the following matrix $\Phi$ 
\begin{align}
\psi = \begin{pmatrix}
 \psi_1 \\ \psi_2 \\ \psi_3 \\ \psi_4
 \end{pmatrix}
\Longleftrightarrow
\Phi = \begin{pmatrix}
 \psi_1 + \psi_3 & - \psi_2^{*} + \psi_4^* \\
 \psi_2 +\psi_4 & \psi_1^* -\psi_3^* 
 \end{pmatrix}.
\end{align} 
Moreover, one should note the following 
\begin{align}
\bar{ \Phi }= \Gamma\Phi^T\Gamma^\dagger= \begin{pmatrix}
 \psi_1^* -\psi_3^*& \psi_2^{*} - \psi_4^* \\
 -\psi_2 - \psi_4 & \psi_1 + \psi_3 
 \end{pmatrix},
\end{align} 
where the superscript $T$ is transposition and $\Gamma=-i\sigma_2$ is the quaternion operator equivalent to a complex conjugation, in terms of which we can write
\begin{align}
\Theta= \begin{pmatrix}
 \Phi& 0 \\
 0 & \bar{\Phi}^\dagger 
 \end{pmatrix}
\end{align}
as a full $4\times4$ matrix.
One should notice the following
\begin{align}\label{APStoSTA}
U\Psi U^\dagger=\Theta
\end{align}
showing that the complete $4\times4$ Baylis form is just the Hestenes matrix form up to a unitary transformation.
The specific unitary transformation is nothing else but the transformation that performs the passage to the chiral (Weyl) representation, that is, $U=\frac{1}{\sqrt{2}}\left(1+\gamma^5\gamma_0\right)$.

The matrix spinor (\ref{APStoSTA}) can be written in the following general form
\begin{align}\label{diagonalPsi}
\Theta&=\begin{pmatrix}
 Q & 0 \\
 0 & \sigma_2 (Q^\dagger)^T\sigma_2
 \end{pmatrix},\nonumber\\
 Q&=r^0-i \mathfrak{s}^0 -i(r^k-i\mathfrak{s}^k)\sigma_k.
\end{align}
The determinant of $\Psi$ is
\begin{eqnarray}
\nonumber
&\Det[\Psi]=\left(r_\mu r^\mu-\mathfrak{s}_\mu \mathfrak{s}^\mu-2ir^\mu \mathfrak{s}_\mu\right)\cdot\\
&\cdot\left(r_\mu r^\mu-\mathfrak{s}_\mu \mathfrak{s}^\mu+2ir^\mu \mathfrak{s}_\mu\right)
\end{eqnarray}
and is non-singular for $r_\mu r^\mu- \mathfrak{s}_\mu \mathfrak{s}^\mu$ and $r^\mu \mathfrak{s}_\mu$ not simultaneously zero. We need these conditions to ensure that $\Psi$ be invertible, as needed for the CRDI procedure. 

Calling $Q=SV$, with $S=\sqrt{r_\mu r^\mu-\mathfrak{s}_\mu \mathfrak{s}^\mu-2ir^\mu \mathfrak{s}_\mu}$ and $\Det[V]=1$ we have that the function $S$ can also be expressed as $S=\sqrt{\rho}e^{i\beta/2}$, where
\begin{align}
\sqrt{\rho}&=\left[((r_\mu r^\mu-\mathfrak{s}_\mu \mathfrak{s}^\mu)^2+4(r^\mu \mathfrak{s}_\mu)^2\right]^{1/2},\label{rho1}\\
\beta&=\pm\arctan\left(\frac{2r^\mu \mathfrak{s}_\mu}{r_\mu r^\mu-\mathfrak{s}_\mu \mathfrak{s}^\mu}\right)\label{beta}.
\end{align}
The minus (plus) sign is for $Q$ ( $\sigma_2 (Q^\dagger)^T\sigma_2$).
Finally, we have
\begin{align}\label{diagonalPsi3}
\Theta&=\sqrt{\rho}\begin{pmatrix}
 e^{-i\beta/2}V & 0 \\
 0 & e^{i\beta/2}\sigma_2 (V^\dagger)^T\sigma_2
 \end{pmatrix}.
 \end{align}
By construction $V$ is a unimodular $2\times2$ complex matrix forming the group $SL(2,\mathbb{C})$, which is the complex $3$-dimensional manifold having $6$ degrees of freedom associated to the parameters of boosts and rotations.

It is instructive to check how this unitary transformation affects equation (\ref{DiracS2})
\begin{equation}
U\psi=\Theta u_1,\quad Uu_1=\frac{1}{\sqrt{2}}\begin{pmatrix}
 1 \\ 0 \\ 1 \\ 0
 \end{pmatrix}.
\end{equation}
Given that $U$ is the unitary transformation connecting the standard representation to the chiral representation, $\Psi$ is the $4\times4$ double cover of the Lorentz group.
It then follows that the matrix spinor $\Psi$ is writable as
\begin{align}
\Psi &=\sqrt{\rho} \exp \left(\boldsymbol{i} \beta/2 \right)\mathcal{R},\\
\mathcal{R}&=U^\dagger\begin{pmatrix}
 V & 0 \\
 0 & \sigma_2 (V^\dagger)^T\sigma_2
 \end{pmatrix}U
\end{align}
with $\mathcal{R}$ a unimodular $4\times4$ complex matrix corresponding to the general Lorentz transformations. 
It is noteworthy that the scalar function $\beta$ (known as Yvon-Takabayashi (YT) angle \cite{takabayasi1957relativistic,yvon1940equations}) is believed to be directly related to antiparticles (see, for instance, \cite{campos2020reply}): specifically, its being zero corresponds to a pure particle state, its being $\pi$ corresponds to a pure antiparticle state, with any value in between corresponding to mixed states \cite{Fabbri:2020ypd}. The mathematical meaning of $\beta$ is, however, very straightforward. It simply describes the dynamics of the internal degrees of freedom of the electron, for instance the spin-orbit coupling. This point will become clear in the examples presented below.
\section{The Manifestly Covariant Relativistic Dynamical Inversion}
\label{CRDI}
Having presented the Hestenes form of the Dirac equation, we are now ready to move on so to make this formalism manifestly covariant \cite{Fabbri:2020ypd}. The ensuing Relativistic Dynamical Inversion technique will therefore be made into the Manifestly Covariant Relativistic Dynamical Inversion technique.
\subsection{General treatment: manifestly covariant Hestenes formalism}\label{HestenesCovIntro}
In what we have done in the previous two sections, we have introduced the Hestenes formalism in standard representation and showed that in the chiral representation it is equivalent to the Baylis formalism; chiral representations are important because they keep the irreducible parts separated. 
Nevertheless, at least in $4$ dimensions, all representations are unitarily equivalent. The same could be said for all possible systems of coordinates used to write the Dirac spinor and the Dirac equation. Just the same, it would be important to have a formalism that is covariant in a manifest way for two reasons: the first is that the passage from cartesian to any other system of coordinates can be done straightforwardly; the second, and most important, is that only in a clearly covariant form can gravity be included. In the following we are then going to consider only the generally covariant form of the Dirac spinor field theory, but employing it to the Hestenes formalism so to keep the advantages of RDI.

In the most general form, complex Lorentz transformations are given in terms of Clifford matrices $\gamma_{a}$ such that (\ref{1}) holds. Then we can define
\begin{eqnarray}
&\frac{1}{4}\left[\gamma_{a},\!\gamma_{b}\right]
\!=\!\sigma_{ab}
\end{eqnarray}
where $\sigma_{ab}$ also verify
\begin{eqnarray}
&2i\sigma_{ab}\!=\!\varepsilon_{abcd}\gamma^{5}\sigma^{cd}
\end{eqnarray}
implicitly defining the $\gamma^{5}$ matrix in terms of the completely antisymmetric pseudo-tensor. We can see that
\begin{eqnarray}
&\gamma_{i}\gamma_{j}\gamma_{k}
\!=\!\gamma_{i}\eta_{jk}-\gamma_{j}\eta_{ik}
\!+\!\gamma_{k}\eta_{ij}
\!+\!i\varepsilon_{ijkq}\gamma^{5}\gamma^{q}
\end{eqnarray}
from which it is possible to get
\begin{eqnarray}
&\{\gamma_{a},\sigma_{bc}\}
=i\varepsilon_{abcd}\gamma^{5}\gamma^{d}\label{anticommgamma}\\
&[\gamma_{a},\sigma_{bc}]
=\eta_{ab}\gamma_{c}\!-\!\eta_{ac}\gamma_{b}\label{commgamma}
\end{eqnarray}
and
\begin{eqnarray}
&\{\sigma_{ab},\sigma_{cd}\}
=\frac{1}{2}[(\eta_{ad}\eta_{bc}\!-\!\eta_{ac}\eta_{bd})\boldsymbol{1}
\!+\!i\varepsilon_{abcd}\gamma^{5}]\label{anticommsigma}\\
&[\sigma_{ab},\sigma_{cd}]
=\eta_{ad}\sigma_{bc}\!-\!\eta_{ac}\sigma_{bd}
\!+\!\eta_{bc}\sigma_{ad}\!-\!\eta_{bd}\sigma_{ac}\label{commsigma}
\end{eqnarray}
are all valid as geometric identities. This last relationship in particular tells us that the $\sigma_{ab}$ matrices are the generators of the Lorentz algebra, so that with parameters $\theta_{ij}\!=\!-\theta_{ji}$ we can write 
\begin{eqnarray}
&\boldsymbol{\Lambda}\!=\!e^{\frac{1}{2}\theta_{ab}\sigma^{ab}}
\end{eqnarray}
as Lorentz transformations. To make them explicit, we define the following quantities
\begin{eqnarray}
a\!=\!-\frac{1}{8}\theta_{ij}\theta^{ij}\\
b\!=\!\frac{1}{16}\theta_{ij}\theta_{ab}\varepsilon^{ijab}
\end{eqnarray}
and then
\begin{eqnarray}
2x^{2}\!=\!a\!+\!\sqrt{a^{2}\!+\!b^{2}}\\
2y^{2}\!=\!-a\!+\!\sqrt{a^{2}\!+\!b^{2}}
\end{eqnarray}
so to introduce the parameters
\begin{eqnarray}
&\cos{y}\cosh{x}\!=\!X\\
&\sin{y}\sinh{x}\!=\!Y\\
\nonumber
&\left(\frac{x\sinh{x}\cos{y}
+y\sin{y}\cosh{x}}{x^{2}+y^{2}}\right)\theta^{ab}+\\
&+\left(\frac{x\cosh{x}\sin{y}
-y\cos{y}\sinh{x}}{x^{2}+y^{2}}\right)\!\frac{1}{2}\theta_{ij}\varepsilon^{ijab}\!=\!Z^{ab}
\end{eqnarray}
which verify
\begin{eqnarray}
X^{2}\!-\!Y^{2}\!+\!\frac{1}{8}Z^{ab}Z_{ab}\!=\!1\label{prop1}\\
2XY\!-\!\frac{1}{16}Z^{ij}Z^{ab}\varepsilon_{ijab}\!=\!0\label{prop2}
\end{eqnarray}
in general. Using (\ref{anticommsigma}) one can prove that
\begin{eqnarray}
\boldsymbol{\Lambda}\!=\!
X\boldsymbol{1}\!+\!Yi\gamma^{5}+\frac{1}{2}Z^{ab}\sigma_{ab}
\label{lambda}
\end{eqnarray}
in the most compact way. One might also be convinced of this result by considering single rotations or boosts and verify by direct inspection that (\ref{lambda}) reduces to the known forms. The inverse is
\begin{eqnarray}
&\boldsymbol{\Lambda}^{-1}\!=\!e^{-\frac{1}{2}\theta_{ab}\sigma^{ab}}
\end{eqnarray}
written explicitly as
\begin{eqnarray}
\boldsymbol{\Lambda}^{-1}\!=\!X\boldsymbol{1}\!+\!Yi\gamma^{5}
\!-\!\frac{1}{2}Z^{ab}\sigma_{ab}
\end{eqnarray}
as clear after using relations (\ref{prop1}-\ref{prop2}) given above.
With this transformation we can define spinor fields as what transforms according to 
\begin{eqnarray}
&\psi\!\rightarrow\!\boldsymbol{\Lambda}\psi
\end{eqnarray}
in the most general case.
Notice that we have
\begin{eqnarray}
(\Lambda)^{a}_{\phantom{a}b}\boldsymbol{\Lambda}\gamma^{b}\boldsymbol{\Lambda}^{-1}\!=\!\gamma^{a}\label{constgamma}
\end{eqnarray}
where $(\Lambda)^{a}_{\phantom{a}b}$ such that $(\Lambda)^{a}_{\phantom{a}k}(\Lambda)^{b}_{\phantom{b}j}\eta^{kj}\!=\!\eta^{ab}$ is a transformation that belongs to the $\mathrm{SO(1,3)}$ group and that is the real representation of the Lorentz transformation.
For future convenience, let us introduce
\begin{eqnarray}
\nonumber
&(\partial_{\mu}XZ^{ab}-X\partial_{\mu}Z^{ab})
+\frac{1}{2}(\partial_{\mu}YZ_{ij}-Y\partial_{\mu}Z_{ij})\varepsilon^{ijab}+\\
&+\partial_{\mu}Z^{ak}Z^{b}_{\phantom{b}k}\!=\!-\partial_{\mu}\zeta^{ab}
\end{eqnarray}
in terms of which we can write
\begin{eqnarray}
\boldsymbol{\Lambda}^{-1}\partial_{\mu}\boldsymbol{\Lambda}
\!=\!\frac{1}{2}\partial_{\mu}\zeta_{ab}\sigma^{ab}
\end{eqnarray}
with (\ref{prop1}-\ref{prop2}) used throughout computations.

Given that the Lorentz transformations can have parameters that are local, we must expect some form of gauge potential and a spinor covariant derivative. By defining the gauge potential as the spinorial connection $\boldsymbol{\Omega}_{\mu}$ transforming as
\begin{eqnarray}
\boldsymbol{\Omega}_{\mu}\!\rightarrow\!\boldsymbol{\Lambda}\left(\boldsymbol{\Omega}_{\mu}
-\boldsymbol{\Lambda}^{-1}\partial_{\mu}\boldsymbol{\Lambda}\right)\boldsymbol{\Lambda}^{-1}
\label{spinconn}
\end{eqnarray}
it is easy to prove that the object
\begin{eqnarray}
&\boldsymbol{\nabla}_{\mu}\psi\!=\!\partial_{\mu}\psi\!+\!\boldsymbol{\Omega}_{\mu}\psi
\end{eqnarray}
transforms as a covariant derivative of the spinor.
Hence, the conditions $\boldsymbol{\nabla}_{\mu}\gamma_{\nu}\!=\!0$ can be expanded into
\begin{eqnarray}
&\boldsymbol{\Omega}_{\mu}\!=\!\frac{1}{2}\Omega_{ij\mu}\sigma^{ij}
\label{spinconndec}
\end{eqnarray}
where
\begin{align*}
\Omega^i_{j\mu}&=e^\nu_je^i_\sigma(\Lambda^\sigma_{\nu\mu}-e^\sigma_a\partial_\mu e^a_\nu)\\
\Lambda^\sigma_{\alpha\nu}&=\frac{g^{\sigma\rho}}{2}(\partial_\alpha g_{\rho\nu}+\partial_\nu g_{\alpha\rho}-\partial_\rho g_{\alpha\nu})
\end{align*}
as the most general (non-conformally invariant) decomposition of the spinor connection. With this the final task described in Sec. \ref{GenDesc} is taken care of. The above discussion is a general introduction to spinor fields and their covariant derivatives. We will now try to see how Hestenes matrix spinors fit into this scheme.

To begin our analysis, let us consider the Hestenes matrix spinor given in (\ref{HestenesG}). The first column is of course the spinor $\psi$ itself. The third column is $\gamma^{5}\psi$ in the standard representation (\ref{gamma5}). As for the second and fourth columns, their meaning may look more cumbersome, but in fact they are merely $\gamma^{5}i\gamma^{2}\psi^{*}$ and $i\gamma^{2}\psi^{*}$ still in the standard representation. That is, any column is obtained from the first after applying the discrete transformations 
\begin{eqnarray}
&\psi\!\rightarrow\!\gamma^{5}\psi\\
&\psi\!\rightarrow\!i\gamma^{2}\psi^{*}
\end{eqnarray}
or combinations thereof. There is a straightforward physical meaning for this: Starting from the first column describing a matter state of defined eigen-spin, the third is just the corresponding antimatter state of same eigen-spin. The second and fourth columns are then the matter and antimatter states of inverted eigen-spin. It is rather important to notice two things: first, despite having deduced everything from the Hestenes formalism given in standard representation, if we were to write the spinor (\ref{HestenesG}) in the form (note that each entrance separated by a | corresponds to a column matrix)
\begin{eqnarray}
&\Psi=\begin{pmatrix}
 \psi &|& \gamma^{5}i\gamma^{2}\psi^{*} 
 &|& \gamma^{5}\psi 
 &|& i\gamma^{2}\psi^{*}
 \end{pmatrix}
\end{eqnarray}
this way of writing the spinor would be representation-independent; second, because we have $\gamma^{2}(\gamma^{\mu})^{*}\gamma^{2}=\gamma^{\mu}$ and $[\gamma^{5},\sigma_{ij}]\!=\!0$, it is easy to prove the following
\begin{eqnarray}
&\gamma^{5}\psi\!\rightarrow\!
\boldsymbol{\Lambda}\gamma^{5}\psi\\
&i\gamma^{2}\psi^{*}\!\rightarrow\!
\boldsymbol{\Lambda}i\gamma^{2}\psi^{*}
\end{eqnarray}
from which one can conclude that
\begin{eqnarray}
&\Psi\!\rightarrow\!\boldsymbol{\Lambda}\Psi
\end{eqnarray}
in general. In polar form (\ref{GeneralMatrixPsi}), because $\rho$ and $\beta$ are real scalars, the transformation law is inherited entirely by $\mathcal{R}$ in the form $\mathcal{R}\!\rightarrow\!\boldsymbol{\Lambda}\mathcal{R}$ showing that this is just what we have for Lorentz transformations. Therefore, the covariant derivative of $\Psi$ is the same as $\psi$ and
\begin{eqnarray}
&\boldsymbol{\nabla}_{\mu}\Psi\!=\!\partial_{\mu}\Psi\!+\!\boldsymbol{\Omega}_{\mu}\Psi
\end{eqnarray}
in the most general situation. From the polar form we have
\begin{eqnarray}
&\!\!\!\!\!\!\!\!\boldsymbol{\nabla}_{\mu}\Psi=\left[\frac{1}{2}(\partial_{\mu}\ln{\rho}
\!+\!\boldsymbol{i}\partial_{\mu}\beta)
\!+\!(\boldsymbol{\Omega}_{\mu}
\!-\!\mathcal{R}\partial_{\mu}\mathcal{R}^{-1})\right]\Psi
\end{eqnarray}
where the first parenthesis contains the covariant derivatives of the two real scalars and we have an additional parenthesis whose meaning is still not obvious. However, we can prove that it is perfectly covariant. To see this, consider that as mentioned above $\mathcal{R}\!\rightarrow\!\boldsymbol{\Lambda}\mathcal{R}$ and combine it with (\ref{spinconn}); putting things together yields
\begin{eqnarray}
\nonumber
&\boldsymbol{\Omega}_{\mu}\!-\!\mathcal{R}\partial_{\mu}\mathcal{R}^{-1}\!\rightarrow\!
\boldsymbol{\Lambda}\left(\boldsymbol{\Omega}_{\mu}
-\boldsymbol{\Lambda}^{-1}\partial_{\mu}\boldsymbol{\Lambda}\right)\boldsymbol{\Lambda}^{-1}-\\
\nonumber
&-\boldsymbol{\Lambda}\mathcal{R}\partial_{\mu}(\mathcal{R}^{-1}\boldsymbol{\Lambda}^{-1})
=\boldsymbol{\Lambda}\boldsymbol{\Omega}_{\mu}\boldsymbol{\Lambda}^{-1}-\\
\nonumber
&-\partial_{\mu}\boldsymbol{\Lambda}\boldsymbol{\Lambda}^{-1}
-\boldsymbol{\Lambda}\mathcal{R}\partial_{\mu}\mathcal{R}^{-1}\boldsymbol{\Lambda}^{-1}-\\
\nonumber
&-\boldsymbol{\Lambda}\mathcal{R}\mathcal{R}^{-1}\partial_{\mu}\boldsymbol{\Lambda}^{-1}
=\boldsymbol{\Lambda}\boldsymbol{\Omega}_{\mu}\boldsymbol{\Lambda}^{-1}-\\
&-\boldsymbol{\Lambda}\mathcal{R}\partial_{\mu}\mathcal{R}^{-1}\boldsymbol{\Lambda}^{-1}\!=\!
\boldsymbol{\Lambda}\left(\boldsymbol{\Omega}_{\mu}
\!-\!\mathcal{R}\partial_{\mu}\mathcal{R}^{-1}\right)\boldsymbol{\Lambda}^{-1}
\end{eqnarray}
as transformation law of the $\boldsymbol{\Omega}_{\mu}-\mathcal{R}\partial_{\mu}\mathcal{R}^{-1}$ object and therefore demonstrating its manifest covariance. Hence we may define
\begin{eqnarray}
&\boldsymbol{\Omega}_{\mu}\!-\!\mathcal{R}\partial_{\mu}\mathcal{R}^{-1}\!=\!-\boldsymbol{R}_{\mu}
\end{eqnarray}
in terms of which 
\begin{eqnarray}
&\!\!\!\!\boldsymbol{\nabla}_{\mu}\Psi=\left[\frac{1}{2}(\nabla_{\mu}\ln{\rho}
\!+\!\boldsymbol{i}\nabla_{\mu}\beta)\!-\!\boldsymbol{R}_{\mu}\right]\Psi
\end{eqnarray}
now clearly manifestly covariant in each term separately.

As a further step, we will see how $\boldsymbol{R}_{\mu}$ decomposes in terms of simpler expressions. Because in general we have
\begin{eqnarray}
&\mathcal{R}\partial_{\mu}\mathcal{R}^{-1}
\!=\!\frac{1}{2}\partial_{\mu}\xi^{ab}\sigma_{ab}
\end{eqnarray}
for some $\xi^{ab}$ and given (\ref{spinconndec}) then we can define
\begin{eqnarray}
&\partial_{\mu}\xi_{ij}\!-\!\Omega_{ij\mu}\!\equiv\!R_{ij\mu}\label{R}
\end{eqnarray}
so that
\begin{eqnarray}
&\boldsymbol{R}_{\mu}\!=\!\frac{1}{2}R_{ij\mu}\sigma^{ij}
\end{eqnarray}
with $R_{ij\mu}$ real tensorial quantities. In fact, as
\begin{eqnarray}
&\frac{1}{2}R_{ij\mu}\sigma^{ij}\!\rightarrow\!\boldsymbol{\Lambda}
\left(\frac{1}{2}R_{ij\mu}\sigma^{ij}\right)\boldsymbol{\Lambda}^{-1}
\end{eqnarray}
the linear independence of the sigmas and (\ref{constgamma}) give
\begin{eqnarray}
&R_{ab\mu}\!\rightarrow\!R_{ij\mu}(\Lambda^{-1})^{i}_{a}(\Lambda^{-1})^{j}_{b}
\end{eqnarray}
showing that $R_{ij\mu}$ has the transformation law of a real tensor. Moreover, the velocity and spin vectors, (\ref{velocity1}) and (\ref{spin1}), satisfy the following geometrical identities
\begin{align}\label{GeosvId}
\nabla_\mu v_i=R_{ij\mu}v^j,\quad \nabla_\mu s_i=R_{ij\mu}s^j.
\end{align}
The spinorial covariant derivative is then
\begin{eqnarray}
&\!\!\!\!\!\!\!\!\boldsymbol{\nabla}_{\mu}\Psi\!=\!(\nabla_{\mu}\ln{\sqrt{\rho}}
\!+\!i\gamma^{5}\nabla_{\mu}\beta/2
\!-\!\frac{1}{2}R_{ij\mu}\sigma^{ij})\Psi
\label{spincovder}
\end{eqnarray}
in the most general circumstance. In fact, (\ref{spincovder}) is valid in any coordinate system, including curved space-times.
\subsection{Introduction of Torsion}
Because we are dealing with generic space-times in which gravity can be present, it may be instructive, for the sake of completeness, to allow also torsion. The Dirac matter field equations in this case are \cite{G}
\begin{eqnarray}
&i\gamma^{\mu}\boldsymbol{\nabla}_{\mu}\psi
\!-\!XW_{\mu}\gamma^{\mu}\gamma^{5}\psi\!-\!m\psi\!=\!0
\label{D}
\end{eqnarray}
with $W_{\mu}$ torsion and $X$ torsion-spin coupling constant.
These equations can be written in the Hestenes formalism by employing the polar form and keeping track of the transformation of the spinor under $\psi\!\rightarrow\!\gamma^{5}\psi$ and $\psi\!\rightarrow\!i\gamma^{2}\psi^{*}$; eventually it is easy to verify that
\begin{eqnarray}
\nonumber
&i\gamma^{\mu}(\nabla_{\mu}\ln{\sqrt{\rho}}
\!+\!i\gamma^{5}\nabla_{\mu}\beta/2)\Psi
\!-\!\frac{i}{2}R_{ij\mu}\gamma^{\mu}\sigma^{ij}\Psi-\\
&-XW_{\mu}\gamma^{\mu}\gamma^{5}\Psi
\!-\!m\Psi\gamma^{3}\gamma^{5}\!=\!0
\label{DH}
\end{eqnarray}
as the Dirac equation for the Hestenes spinor. Equation (\ref{DH}) is manifestly covariant as is straightforward to see.

\subsection{Electromagnetic Interaction and Inversion of the Gauge Potential}
Having written the Dirac equation in curvilinear coordinates for curved space-times in presence of torsion, the only missing interaction for single particle systems is electrodynamics. In order to add it we will employ the concept of gauge covariance. When gauge covariance is present, it simply means that complex objects transform according to an additional phase $e^{iq\xi}$. One may then wonder: What is the effect of such a phase shift on the Hestenes spinor? In order to answer this question, we have to consider that the discrete transformations $\psi\!\rightarrow\!\gamma^{5}\psi$ and $\psi\!\rightarrow\!i\gamma^{2}\psi^{*}$ lead to
\begin{eqnarray}
&\gamma^{5}\psi\!\rightarrow\!e^{iq\xi}\gamma^{5}\psi\\
&i\gamma^{2}\psi^{*}\!\rightarrow\!e^{-iq\xi}i\gamma^{2}\psi^{*}
\end{eqnarray}
where now the phase is no longer the same for the second and fourth columns. As a consequence, the full transformation for the gauge phase would be 
\begin{eqnarray}
&\Psi\!\rightarrow\!\Psi\exp{(-q\xi\gamma^{1}\gamma^{2})}
\end{eqnarray}
which can be easily checked. We notice that the phase ultimately acts as if it were a rotation around the third axis.

The covariant derivatives has to be up-dated with the gauge potential which transforms as
\begin{eqnarray}
&A_{\mu}\!\rightarrow\!A_{\mu}\!-\!\partial_{\mu}\xi
\end{eqnarray}
ensuring that
\begin{eqnarray}
&\boldsymbol{\nabla}_{\mu}\Psi\!=\!\partial_{\mu}\Psi
\!-\!qA_{\mu}\Psi\gamma^{1}\gamma^{2}
\end{eqnarray}
is the gauge covariant derivative of the matrix spinor.
Thence, the matrix spinor in polar form now reads
\begin{align}
\Psi=\sqrt{\rho} \exp\left(\boldsymbol{i} \beta/2 \right)\mathcal{R}
\exp{(q\xi\gamma^{1}\gamma^{2})}
\end{align}
so that upon introduction of the object
\begin{eqnarray}
&q(\partial_{\mu}\xi\!-\!A_{\mu})\!\equiv\!P_{\mu}\label{P}
\end{eqnarray}
proven to be a real vector, one can write
\begin{eqnarray}
&\nonumber
\!\!\!\!\!\!\!\!\boldsymbol{\nabla}_{\mu}\Psi\!=\!(\nabla_{\mu}\ln{\sqrt{\rho}}
\!+\!i\gamma^{5}\nabla_{\mu}\beta/2-\\
&-\frac{1}{2}R_{ij\mu}\sigma^{ij})\Psi
+P_{\mu}\Psi\gamma^{1}\gamma^{2}
\label{Der}
\end{eqnarray}
as can be easily checked.

Finally, the Dirac equations are
\begin{align}\label{CoVHestenes}
&i\gamma^{\mu}(\nabla_{\mu}\ln{\sqrt{\rho}}
\!+\!i\gamma^{5}\nabla_{\mu}\beta/2)\Psi
\!-\!\frac{i}{2}R_{ij\mu}\gamma^{\mu}\sigma^{ij}\Psi\nonumber\\
&+iP_{\mu}\gamma^{\mu}\Psi\gamma^{1}\gamma^{2}
\!-\!XW_{\mu}\gamma^{\mu}\gamma^{5}\Psi
\!-\!m\Psi\gamma^{3}\gamma^{5}\!=\!0
\end{align}
which are the most general equations, despite the explicitly appearance of $\gamma_a$ in it, which are determined only within a proper Lorentz transformation. So it cannot be overemphasised that the vectors $\gamma_a$ need not be associated a priori with any coordinate systems, but they are simply a set of arbitrarily chosen orthonormal vectors, much in the same way in which $\gamma_2$ is used to define the procedure of charge conjugation without implying any dependence on the second axis of the coordinate system.

The advantage of writing (\ref{CoVHestenes}) is straightforward: because the spinor is in the form of an invertible matrix, it is immediate to invert the Dirac equation with respect to the electrodynamic potential. Then, solutions of the Maxwell equations can be obtained very easily. Before describing the inversion, let us massage a bit (\ref{CoVHestenes}), so that one can more easily compare it with (\ref{Dirac-Hestenes-Eq5}), by multiplying it to the right with $\gamma^2\gamma^1$. Hence, after collecting all the terms we end up with
\begin{eqnarray}
\nonumber
&\gamma^{\mu}\left(\nabla_{\mu}\ln{\sqrt{\rho}}
\!-\!\boldsymbol{i}(\nabla_{\mu}\beta/2+XW_\mu)-\frac{1}{2}R_{ij\mu}\sigma^{ij}\right)\Psi\gamma^2\gamma^1\\
&+P_{\mu}\gamma^{\mu}\Psi
-\!m\Psi\gamma^{0}\!=\!0.
\end{eqnarray}
The inversion is then
\begin{eqnarray}\label{inversionCov}
\nonumber
&[\gamma^{\mu}\left(\nabla_{\mu}\ln{\sqrt{\rho}}
\!-\!\boldsymbol{i}(\nabla_{\mu}\beta/2+XW_\mu)-\frac{1}{2}R_{ij\mu}\sigma^{ij}\right)\Psi\gamma^2\gamma^1\\
&-\!m\Psi\gamma^{0}]\Psi^{-1}\!=-\!\gamma^{\mu}P_{\mu}
\end{eqnarray}
showing that it is possible to write $P_{\mu}$ in terms of something that does not contain $P_{\mu}$ itself. Because $P_{\mu}$ is the only instance containing $A_{\mu}$, electrodynamics is inverted. Since we will not be considering torsion in this paper, hereafter we choose $X=0$ in (\ref{CoVHestenes}) and (\ref{inversionCov}).

By following the procedure of \cite{Fabbri:2019tad} one can see that
\begin{eqnarray}\label{VecPotEq}
\nonumber
&P^{\eta}\!=\!mc\cos{\beta}v^\eta\!+\!\frac{\hbar}{2}[(\partial_\mu\beta\!+\!B_\mu)v^\mu s^\eta-\\
&-(\partial_\mu\beta\!+\!B_\mu)s^\mu v^\eta
\!-\!(\partial_\mu\ln{\rho}\!+\!R_\mu)s_\alpha v_\nu\varepsilon^{\mu\alpha\nu\eta}]
\end{eqnarray}
where $R_\mu=R_{\mu a}^{\,\,\,\,\,\,a}$ and $B_\mu=\frac{1}{2}\varepsilon_{\mu\alpha\nu\eta}R^{\alpha\nu\eta}$. It is noteworthy that in order to calculate the vector potential from (\ref{VecPotEq}), one needs to get the functions $\beta$, $s_\mu$, $v_\mu$, $B_\mu$ and $R_\mu$ using the matrix spinor $\Psi$. We notice that according to (\ref{VecPotEq}) $P^{\eta}$ is a real function, so taking (\ref{P}) written as $A_{\mu}\!\equiv\!\partial_{\mu}\xi-P_{\mu}/q$ shows that $A^{\eta}$ is also real, and therefore $A^{\eta}$ describes a real electrodynamic potential, as expected. This seems to indicate that $P^{\eta}$ is the momentum of the particle. This can definitively be accepted by considering that in the case of plane waves (\ref{Der}) reduces to
\begin{eqnarray}
\boldsymbol{\nabla}_{\mu}\Psi\!=\!P_{\mu}\Psi\gamma^{1}\gamma^{2}
\end{eqnarray}
which is equivalent to
\begin{eqnarray}
\boldsymbol{\nabla}_{\mu}\psi\!=\!-iP_{\mu}\psi
\end{eqnarray}
as can be seen in \cite{Moulin:2021rup}. This is precisely the definition of the quantum mechanical momentum. A final confirmation comes form its macroscopic limit, taken when the spin is approximated to zero. This implies that $s^\eta$ as well as $\beta$ vanish \cite{Fabbri:2020ypd}. In this case $P^{\eta}\!\rightarrow\!mcv^\eta$ which is precisely the kinematic momentum. This also shows that the full momentum is given by the simplest kinematic momentum with two corrections. One is the multiplication by $\cos{\beta}$ and it accounts for the internal dynamics \cite{Fabbri:2020ypd}, while the other is the addition of a term that is linear in $\hbar$ and as such accounts for quantum mechanical corrections. In fact, it can be proven that these corrections are precisely the quantum potential of the de Broglie-Bohm theory extended to the relativistic case with spin \cite{Fabbri:2022gio}.

Also, from the Dirac equation we extract the following constraints on the velocity $\rho v\!\!\!/$ and spin $\rho s\!\!\!/$ vector densities
\begin{align}
&\nabla_\mu(\rho v^\mu)=0\label{continuityEq}\\
&\nabla_\mu(\rho s^\mu)=-\frac{2mc}{\hbar}\rho\sin\beta.\label{continuityEqSpin}
\end{align}
Equations (\ref{VecPotEq}), (\ref{continuityEq}) and  (\ref{continuityEqSpin}) correspond to the covariant generalization of (\ref{scalarpart}), (\ref{vectorpart}) and (\ref{constraints}). Note that, based on the discussion of Sec. \ref{RDI}, an important feature of CRDI is that it can be straightforwardly connect with the quaternion formalism of G\"ursey (see also \cite{gursey1956correspondence}).

In what follows we will apply (\ref{VecPotEq}) ,(\ref{continuityEq}) and (\ref{continuityEqSpin}) to construct a family of analytical solutions to the Dirac equation in spherical coordinates in order to illustrate the power of CRDI.
\section{Illustrations of the method}
\label{illustration}
In this section, we start with the Dirac representation of the gamma matrices in cartesian coordinates and will transform them to spherical coordinates using tetrads.
Moreover, in addition to presenting an example for the method, we will also choose the example to be representing a specific physical situation that we find extremely intriguing. Namely, the one in which the YT angle is taken to generate a spin-orbit coupling.

\subsection{YT angle and spin-orbit coupling}
As a prototype of the matrix spinor we will consider the structure originally taken in \cite{FabbriandCampos2021,Fabbri:2021weq}, and that is
\begin{align}
\Psi&=\sqrt{\rho}e^{\boldsymbol{i}\frac{\beta}{2}}\mathcal{B}Ue^{-\gamma_2\gamma_1\epsilon t/\hbar},\label{GenSpinor}\\
\mathcal{B}&=e^{\left(-\sin\phi\gamma_1\gamma_0+\cos{\phi}\gamma_2\gamma_0\right)\frac{w}{2}},\\
U&=e^{-\gamma_2\gamma_1\phi/2}e^{-\gamma_3\gamma_1\arctan\left[\tan(\frac{\beta}{2})\tanh(\frac{w}{2})\right]}
 e^{\gamma_2\gamma_1\phi/2}\label{GenU},
\end{align}
where $\phi=\arctan{y/x}$, $\beta=\arctan{g}$, $w=\arctanh{f}$, $g=g(x,y,z)$ and $f=f(x,y,z)$. This form is general in the sense that it contains the module $\rho$ and the chiral angle $\beta$ as two physical degrees of freedom. The other degrees of freedom are the real constant $\epsilon$, the rotation $U$ and the boost $\mathcal{B}$. In analogy to the classical Kepler problem, we consider an electron moving in a circular orbit on the $x-y$ plane; such a feature is described by the boost. Since the electron is in an accelerated frame there is also a precession of the spin vector, which is given by $U$. The specific form of the rotation matrix is a consequence of the conservation of total angular momentum due to its dependence on the rapidity $\tanh(\frac{w}{2})$, which follows from the spin-orbit coupling. Thus, the physically relevant potentials derived from the matrix spinor (\ref{GenSpinor}) are the ones for which the total angular momentum is conserved.

\subsubsection{2D time independent solution}
Here we briefly discuss the relationship of (\ref{GenSpinor}) with the solutions presented in \cite{RDI3} . Let us consider a solution with $g=0$ in (\ref{GenSpinor}) so that there will be no rotation in the $\gamma_3\gamma_1$ plane, only in the $\gamma_2\gamma_1$ plane. The dynamics is thus confined to the $x-y$ plane, with the system possessing cylindrical symmetry, since the boost is also along the $\gamma_2\gamma_0$ direction (i.e., the $y$ direction). The matrix spinor (\ref{GenSpinor}) then becomes
\begin{align}\label{GenSpinor2D}
\Psi&=\sqrt{\rho}e^{-\gamma_2\gamma_1\phi/2}e^{\gamma_2\gamma_0\frac{\arctanh{f}}{2}}e^{\gamma_2\gamma_1\phi/2}e^{-\gamma_2\gamma_1\frac{\epsilon t}{\hbar}}\nonumber\\
&=\sqrt{\rho}e^{\left(-\sin\phi\gamma_1\gamma_0+\cos{\phi}\gamma_2\gamma_0\right)\frac{\arctanh{f}}{2}}e^{-\gamma_2\gamma_1\frac{\epsilon t}{\hbar}}
\end{align}
which is just a boost with rapidity $w=\arctanh{f(r)}$, in which $r=\sqrt{x^2+y^2}$, along the azimuthal direction in the $x-y$ plane together with the rotation around the $\hat{z}$ axis. It is noteworthy that the matrix spinor (\ref{GenSpinor2D}) corresponds to a solution with zero orbital angular momentum $L$. The generalization to include it (i.e., by considering  the corresponding quantum number $l\neq0$) can be done by simply multiplying (\ref{GenSpinor2D}) on the right with a matrix proportional to $(re^{-\gamma_2\gamma_1\phi})^{-l}$ as we have previously described in \cite{RDI3}. Moreover, the matrix spinor (\ref{GenSpinor2D}), with the addition of the orbital angular momentum term, corresponds to the one underlying \textit{all} stationary solutions presented in \cite{RDI3}.

\subsubsection{3D time independent solution}
In this case we work with the matrix spinor (\ref{GenSpinor}) in its full glory. In order to have non-zero YT angle, from (\ref{beta}) we require that $r^\mu \mathfrak{s}_\mu\neq0$ and $r_\mu r^\mu-\mathfrak{s}_\mu \mathfrak{s}^\mu\neq0$. In keeping with this requirement, there is a particular form of $f$ and $g$ that leads to the ground state of the Hydrogen atom; that is
\begin{align}\label{choice1}
f=\frac{\sin\theta}{\sqrt{1+X^2}},\quad g=\frac{\cos{\theta}}{X},
\end{align}
for constant $X=\sqrt{1-Z^2\alpha^2}/Z\alpha$, where $Z$ is the atomic charge and $\alpha$ is the fine structure constant. Note that the function $f$ is nothing but the magnitude of the electron's velocity. In order to generalize this solution, let us consider $X=X(r)$ a general real function of $r$ alone. Moreover, let us make the following choice of density
\begin{align}\label{densityHatom}
\sqrt{\rho}=\kappa\frac{e^{-G/2}}{r}(X^2+|\cos{\theta}|^2)^{1/4},
\end{align}
for some function $G=G(r)$ and where $\kappa$ is a normalization constant.

Substituting in (\ref{GenSpinor}) gives
\begin{align}\label{GenSpinor2}
&\Psi=\sqrt{\rho}e^{\boldsymbol{i}\frac{\beta}{2}}\mathcal{B}Ue^{-\gamma_2\gamma_1(\epsilon t/\hbar-\phi/2)},\\
&\mathcal{B}=e^{\left(-\sin\phi\gamma_1\gamma_0+\cos{\phi}\gamma_2\gamma_0\right)\frac{\arctanh(f)}{2}},\\
&U=e^{-\gamma_2\gamma_1\phi/2}e^{-\gamma_3\gamma_1\Theta/2},\\
&\Theta=\arccos\left(\frac{|\!\sin{\theta}|^2 X
+|\!\cos{\theta}|^2\sqrt{X^2+1}}{\sqrt{|\!\cos{\theta}|^2+X^2}}\right)
\end{align}
from which we extract the spinor
\begin{align}\label{spinorHatom}
\psi=\frac{\kappa e^{-\frac{i\epsilon t}{\hbar}-\frac{G }{2 }}}{\sqrt{2}r}\begin{pmatrix}\sqrt{X+\sqrt{1+X^2}}
 \\ 0\\ i\frac{\cos{\theta}}{\sqrt{X+\sqrt{1+X^2}}}\\ i\frac{e^{i\phi}\sin\theta}{\sqrt{X+\sqrt{1+X^2}}}\end{pmatrix}
\end{align}
that solves the Dirac equation for the vector potential $A_0=-V(r)$ and $\vec{A}=0$ calculated from (\ref{scalarpart}) and (\ref{vectorpart}), as long as the following constraints
\begin{align}
\frac{dX}{dr}&=-2\frac{\sqrt{X^2+1}}{r}\Bigg(r\left[\frac{\epsilon}{\hbar c}+\frac{V}{\hbar}\right]\sqrt{1+X^2}\nonumber\\
&-1-\frac{mc}{\hbar}rX\Bigg)\label{CondX1},\\
\frac{dG}{dr}&=2\left(\frac{mc}{\hbar}\sqrt{1+X^2}-X\left[\frac{\epsilon}{\hbar c}+\frac{V}{\hbar}\right]\right)\label{CondG1}.
\end{align}
are imposed on $X$ and $G$. The constraints imposed on $X$ and $G$ are needed so that the second equation in (\ref{constraints}) is satisfied while the first equation is obeyed regardless.
Before proceeding, let us analyze the definition (\ref{densityHatom}). Such a form of density is not arbitrary; it is given by (\ref{rho1}). There is another way of getting the density that relies solely on the matrix part of (\ref{GenSpinor2}). Consider a general function $\sqrt{\rho}$. By looking at the matrix $e^{\boldsymbol{i}\frac{\beta}{2}}\mathcal{B}Ue^{\gamma_2\gamma_1\phi/2}$ one notes that all of its columns are multiplied by the term $(X^2+|\cos{\theta}|^2)^{-1/4}$. Given that the potential has spherical symmetry, the components of $\psi$ (i.e., the first column of $\Psi$) must be written as the product of a radial only and an angular only functions (e.g., $\zeta(r)\eta(\theta,\phi)$). Since the only term that cannot be written in such way is the overall multiplying function $(X^2+|\cos{\theta}|^2)^{-1/4}$, one must choose $\sqrt{\rho}=\zeta(r)(X^2
+|\cos{\theta}|^2)^{1/4}$ for some real function $\zeta(r)$ which must obey the constraint $\lim_{r\rightarrow\infty}\zeta(r)=0$. The specific form of $\zeta(r)$ chosen here is inspired by the spinor in the ground state of the Hydrogen atom.

The equation for $X$ can be cast into the form of a Riccati differential equation. For instance, upon making the substitution $X=-\csch(2\arctanh(Z(r)))$ we get
\begin{align}
\frac{dZ}{dr}&=\frac{Z(r)^2 \left(c^2 m-c V-\epsilon \right)}{c \hbar }-\frac{c^2 m+c
 V+\epsilon }{c \hbar }\nonumber\\
 &+\frac{2 Z(r)}{r}\label{CondEq1},\\
\frac{dG}{dr}&=\frac{Z(r) \left(c^2m-c V-\epsilon \right)}{c \hbar }+\frac{c^2 m+c
 V+\epsilon }{c \hbar Z(r)}\label{CondEq2}.
\end{align}
Hence a solution of the equation for $Z(r)$ that is guaranteed to exist, should be substituted into the equation for $G$ which is then integrated.

We next move from Cartesian to spherical coordinates.
To do that, we first notice that (\ref{densityHatom}) and (\ref{spinorHatom}) retain their current form in spherical coordinates since the spinors transform as scalars under a change of coordinates. To build the tetrads, we start by considering the metric tensor
\begin{eqnarray}
&\!\!\!\!\!\!\!\!g_{tt}\!=\!1\ \ \ \ g_{rr}\!=\!-1\ \ 
g_{\theta\theta}\!=\!-r^{2}\ \ g_{\phi\phi}\!=\!-r^{2}|\!\sin{\theta}|^{2}
\end{eqnarray}
giving connection
\begin{eqnarray}
&\Lambda^{\theta}_{\theta r}\!=\!\Lambda^{\phi}_{\phi r}\!=\!\frac{1}{r}\\
&\Lambda^{r}_{\theta\theta}\!=\!-r\\
&\Lambda^{r}_{\phi\phi}\!=\!-r|\!\sin{\theta}|^{2}\\
&\Lambda^{\phi}_{\phi\theta}\!=\!\cot{\theta}\\
&\Lambda^{\theta}_{\phi\phi}\!=\!-\cos{\theta}\sin{\theta}
\end{eqnarray}
that are used in the calculation of the components of the spin connection $\boldsymbol{\Omega}_\mu$. Next we use the unitary operator $\mathcal{U}=e^{-\gamma_2\gamma_1\phi/2}e^{-\gamma_1\gamma_3\theta/2}$
in order to construct the gamma matrices on the manifold $\gamma_\nu$ (note that $\gamma_t=\gamma_0$ in this case)
\begin{align}\label{tetradsSp}
&\gamma_r=\mathcal{U}\gamma_3\mathcal{U}^\dagger \ \ \ \ \gamma_\theta=r\mathcal{U}\gamma_1\mathcal{U}^\dagger\ \ \ \ \gamma_\phi=r|\sin\theta|\mathcal{U}\gamma_2\mathcal{U}^\dagger\\
&\gamma^r=\mathcal{U}\gamma^3\mathcal{U}^\dagger \ \ \ \ \gamma^\theta=\frac{1}{r}\mathcal{U}\gamma^1\mathcal{U}^\dagger\ \ \ \ \gamma^\phi=\frac{1}{r|\sin\theta|}\mathcal{U}\gamma^2\mathcal{U}^\dagger\
\end{align}
as can be easily checked. From $e^a_\nu\gamma_a=\gamma_\nu$ we extract the tetrads
\begin{eqnarray}
&e^{0}_{t}=1\\
&e^{1}_{r}\!=\!\cos{\phi}\sin\theta\ \ \ \ e^{2}_r=\sin\phi\sin\theta\\
&e^3_r=\cos{\theta}\\
&e^{1}_{\theta}\!=\!r\cos{\theta}\cos{\phi}\ \ \ \ e^{2}_\theta=r\cos{\theta}\sin\phi\\
&e^{3}_\theta=-r\sin\theta\\
&e^{1}_{\phi}=-r\sin\phi\sin{\theta}\ \ \ \ e^{2}_{\phi}=r\cos{\phi}\sin{\theta}
\end{eqnarray}
with dual
\begin{eqnarray}
&e^{t}_{0}=1\\ 
&e^{r}_{1}=\cos{\phi}\sin\theta\ \ \ \ e^{r}_2=\sin\phi\sin\theta\\
&e^r_3=\cos{\theta}\\
&e^{\theta}_{1}=\frac{1}{r}\cos{\theta}\cos{\phi}\ \ \ \ e^{\theta}_2=\frac{1}{r}\cos{\theta}\sin\phi\\
&e^{\theta}_3=-\frac{1}{r}\sin\theta\\
&e^{\phi}_{1}=-\sin\phi\frac{1}{r\sin{\theta}}\ \ \ \ e^{\phi}_{2}=\cos{\phi}\frac{1}{r\sin{\theta}}.
\end{eqnarray}
It turns out that the above tetrads give a zero spin connection. Also, with the matrix spinor (\ref{GenSpinor}) we calculate the following normalized components of the spin (\ref{spin1}) and velocity (\ref{velocity1}) leading to
\begin{align}
&s^1=\frac{\left(-X+\sqrt{X^2+1}\right)\sin{\theta}\cos{\theta}\cos{\phi}}{\sqrt{|\cos{\theta}|^2+X^2}}\label{s1E}\\
&s^2=\frac{\left(-X+\sqrt{X^2+1}\right)\sin{\theta}\cos{\theta}\sin\phi}{\sqrt{|\cos{\theta}|^2+X^2}}\label{s2E}\\
&s^3=\frac{|\sin{\theta}|^2X+|\cos{\theta}|^2\sqrt{X^2+1}}{\sqrt{|\cos{\theta}|^2+X^2}}\label{s3E}\\
 &v^0=\frac{\sqrt{X^2+1}}{\sqrt{|\cos{\theta}|^2+X^2}}, \,\,\, v^1=-\frac{\sin{\theta}\sin\phi}{\sqrt{|\cos{\theta}|^2+X^2}}\\
 &v^2=\frac{\sin{\theta} \cos{\phi}}{\sqrt{|\cos{\theta}|^2+X^2}}.
\end{align}
where we recall that $X\!=\!X(r)$ everywhere.

Let us pause now, and consider for a moment the components of the spin vector. By making the definition
$$
\mathcal{V}=\frac{\left(-X+\sqrt{X^2+1}\right)\sin{\theta}\cos{\theta}}{\sqrt{|\cos{\theta}|^2+X^2}},
$$
with range $0\leq\mathcal{V}\leq (\sqrt{X^2+1}-X)/\sqrt{2+4X^2}$, the spin components (\ref{s1E}), (\ref{s2E}) and (\ref{s3E}) then take the simple form
\begin{align}\label{sP}
s^1=\mathcal{V}\cos\phi,\,\,s^2=\mathcal{V}\sin\phi,\,\,s^3=\sqrt{1-\mathcal{V}^2}
\end{align}
which, for the case of the H atom (i.e., if $X$ is constant), is just the parametrization of the upper hemisphere of a sphere of radius $(\sqrt{X^2+1}-X)/\sqrt{2+4X^2}$;
the geometrical interpretation of this result is straightforward: considering (\ref{sP}) as a parametric equation with parameters $\phi$ and $\theta$, it can be interpreted that the spin
vector precess, with its tip constrained to move on the upper hemisphere of a sphere whose radius is a function of $X$. Hence, the chosen matrix spinor (\ref{GenSpinor})
successfully describes the motion of an electron whose velocity lies on the $x-y$ plane and whose spin vector wobbles on a surface in which the polar angle lies in the region $0\leq\theta\leq\pi/4$ (this is because $\mathcal{V}$ attains its maximal value for $\theta=\pi/4$). In the general case with $X=X(r)$, the aforementioned geometrical picture does not strictly holds (meaning that while the spin vector still precess, its wobbling will lay on a more general surface).
It is noteworthy that the above interpretation is based on the idea that the spinor is a quantum field, which can be seem as some sort of fluid, the streamlines of which are the electron trajectories in the given electromagnetic field. Hence, it is in this context that the idea of motion is applied, given that everything is time independent. 

With the tetrads one can make the transition from the tangent space to the manifold with $s^\mu=s^ae_a^\mu$ and $v^\mu=v^ae_a^\mu$ giving the following non-zero components
\begin{align}
&s^r=\frac{\cos{\theta} \sqrt{X^2+1}}{\sqrt{|\!\cos{\theta}|^2+X^2}}, \,\, s^\theta=-\frac{\sin{\theta} X}{r \sqrt{|\!\cos{\theta}|^2+X^2}}\\
&v^t=\frac{\sqrt{X^2+1}}{\sqrt{|\!\cos{\theta}|^2+X^2}}, \,\, v^\phi=\frac{1}{r \sqrt{|\!\cos{\theta}|^2+X^2}}.
\end{align}
The $R_{ij\mu}$ tensor is calculated as the solution to the equation
\begin{eqnarray}
&2\mathcal{R}\partial_{\mu}\mathcal{R}^{-1}\!-\!\Omega_{ij\mu}\sigma^{ij}\!=\!R_{ij\mu}\sigma^{ij}
\end{eqnarray}
with $\mathcal{R}\!=\!\mathcal{B}U$ above. 

Let us illustrate here a very powerful property of the proposed method, which is the possibility of easily writing the solution to the Dirac equation \textit{in any} frame of reference. First, one should remember that the solutions to the Dirac equation are written in a frame of reference at rest located at the origin of any chosen coordinate system. For instance, in the case of the Hydrogen atom, this is the frame of the proton (to be precise, it should be the rest frame of the center-of-mass of the proton-electron system, but because the proton is much more massive than the electron, the center-of-mass and the proton are approximately the same). The electron is seen by the observer as undergoing a motion composed of translations and rotations that can easily be inferred from the matrix spinor (\ref{GenSpinor}). Here we show how to change from the aforementioned frame of reference to the rest frame of the electron. The rest frame of the electron is the one for which the spinor takes the form such that the spatial part of the velocity operator is zero. The attentive reader might complain that the change of these two systems might be meaningless because in this transfer we move from one inertial to a non-inertial frame. Such a reader would in principle be right, as in fact there would have to be difficulties in treating non-inertial frames with the standard methods. However, our method here instead is fully covariant, and as such it is naturally equipped to treat non-inertial as well as inertial frames with equal ease, the information about the acceleration of the frame being contained in the spin connection.

In order to change reference frames we first have to calculate the new tetrads. The tetrads are calculated with the following simple formulas
\begin{align}
&e^t_a=\frac{1}{4}\Tr(\mathcal{R}^{-1}\gamma^0\mathcal{R}\gamma_a),\,\, e^\theta_a=\frac{1}{4}\Tr(\mathcal{R}^{-1}\gamma^\theta\mathcal{R}\gamma_a)\\
&e^\phi_a=\frac{1}{4}\Tr(\mathcal{R}^{-1}\gamma^\phi\mathcal{R}\gamma_a),\,\, e^r_a=\frac{1}{4}\Tr(\mathcal{R}^{-1}\gamma^r\mathcal{R}\gamma_a)
\end{align} 
whose components are
\begin{align*}
&e^t_0=\frac{\sqrt{X^2+1}}{\sqrt{|\!\cos{\theta}|^2+X^2}},\,\, e^t_2=\frac{\sin{\theta}}{\sqrt{|\!\cos{\theta}|^2+X^2}}\\
&e^r_1=\frac{\sin{\theta} X}{\sqrt{|\!\cos{\theta}|^2+X^2}},\,\, e^r_3=\cos{\theta} \frac{\sqrt{X^2+1}}{\sqrt{|\!\cos{\theta}|^2+X^2}}\\
&e^\theta_1=\frac{\cos{\theta}}{r} \frac{\sqrt{X^2+1}}{\sqrt{|\!\cos{\theta}|^2+X^2}},\,\, e^\theta_3=-\frac{\sin{\theta} X}{r\sqrt{|\!\cos{\theta}|^2+X^2}}\\
&e^\phi_0=\frac{1}{r \sqrt{|\!\cos{\theta}|^2+X^2}},\,\, e^\phi_2=\frac{1}{r\sin\theta}\frac{\sqrt{X^2+1}}{\sqrt{|\!\cos{\theta}|^2+X^2}}
\end{align*}
as it can be straightforwardly seen.

The electron rest frame is the one for which the spinor-tetrad system gives the velocity $v_{a}\!=\!(1,0,0,0)$ (this can always be done so long as the spinor is non-singular, that is if the matrix spinor has determinant not equal to zero identically in general). In this frame we are also going to pick the spin aligned along the third axis (which can always be done). In this case then, the matrix spinor is simply
\begin{align}\label{RestF}
\Psi_{RF}=\sqrt{\rho}e^{\boldsymbol{i}\frac{\beta}{2}}\boldsymbol{1}e^{-\gamma_2\gamma_1(\epsilon t/\hbar-\phi/2)},
\end{align}
which is just the matrix spinor (\ref{GenSpinor}) in which the replacement $\mathcal{R}\rightarrow\boldsymbol{1}$ is made. Moreover, the components of the spin connection describing the acceleration of the frame are
\begin{align}\label{SpinConnH}
&\Omega_{02r}=-\frac{\sin{\theta} X X'}{\sqrt{X^2+1} \left(|\cos{\theta}|^2+X^2\right)}\\
&\Omega_{13r}=\frac{\sin{\theta} \cos{\theta}X'}{\sqrt{X^2+1} \left(|\cos{\theta}|^2+X^2\right)}\\
&\Omega_{02\theta}=\frac{\cos{\theta} \sqrt{X^2+1}}{|\!\cos{\theta}|^2+X^2}\\
&\Omega_{13\theta}=\frac{X \sqrt{X^2+1}}{|\!\cos{\theta}|^2+X^2}-1\\
&\Omega_{01\phi}=-\frac{\sin{\theta} \left(|\!\sin{\theta}|^2 X+|\!\cos{\theta}|^2\sqrt{X^2+1}\right)}{|\!\cos{\theta}|^2+X^2}\\
&\Omega_{03\phi}=\frac{|\!\sin{\theta}|^2 \cos{\theta} \left(X-\sqrt{X^2+1}\right)}{|\cos{\theta}|^2+X^2}\\
&\Omega_{12\phi}=\frac{|\!\sin{\theta}|^2 X \sqrt{X^2+1}+|\!\cos{\theta}|^2\left(X^2+1\right)}{|\!\cos{\theta}|^2+X^2}\\
&\Omega_{23\phi}=\frac{\sin{\theta} \cos{\theta} \left(X\left(\sqrt{X^2+1}-X\right)-1\right)}{|\!\cos{\theta}|^2+X^2}.
\end{align}
Therefore, the Dirac spinor in the rest frame
\begin{align}\label{Diracrest}
\psi_{RF}&=\Psi_{RF}\begin{pmatrix}
 1 \\ 0 \\ 0 \\ 0
 \end{pmatrix} =\frac{\kappa e^{-\frac{G}{2}-\frac{i t \epsilon }{\hbar }+\frac{i \phi
 }{2}}}{\sqrt{2} r}\nonumber\\
 &\times\begin{pmatrix}
 \sqrt{\sqrt{|\!\cos{\theta}|^2+X^2}+X} \\ 0 \\ \sqrt{\sqrt{|\!\cos{\theta}|^2+X^2}-X} \\ 0
 \end{pmatrix},
 \end{align}
along with the tetrads and the spin connection and the vector potential $A_t=-V(r)$ obeys the Dirac equation
\begin{align}
&i\hbar e^\mu_a\gamma^a\left(\partial_\mu+\frac{1}{2}\Omega_{ij\mu}\sigma^{ij}\right)\psi_{RF}-q\,e^t_a\gamma^a A_t\psi_{RF}\nonumber\\
&-mc\psi_{RF}=0
\end{align}
as long as conditions (\ref{CondX1}) and (\ref{CondG1}) are satisfied.
It is noteworthy that with the choice of functions (\ref{choice1}) if one fixes $\theta=\pi/2$ (i.e., a projection onto the $x-y$ plane), the matrix spinor (\ref{GenSpinor}) takes the same form as (\ref{GenSpinor2D}).

The solution (\ref{Diracrest}) was first presented in \cite{FabbriandCampos2021,Fabbri:2021weq}. The most intriguing property of this spinor is that it is \textit{not} separable,
even though the potential has spherical symmetry. In Ref. \cite{FabbriandCampos2021} it was not explicitly shown, but instead speculated that the non-separability of the Dirac spinor was a consequence of the frame in which the solution was written. In this work we were able to prove the connection between non-separability and reference frame. In fact, if we compare (\ref{Diracrest}) with the spinor (\ref{spinorHatom}), which is the solution to the Dirac equation for the same potential albeit in another frame of reference, we note that variable separability is restored. Hence, we can conclude that the symmetries of equations are not always inherited by their solutions. In fact, here we just proved that such symmetries, and in particular the property of variable separability for the Dirac equation, are \textit{frame dependent}.

The cases just discussed represent physical situations involving the presence of the YT chiral angle and the fact that it is strictly connected to spin-orbit coupling effects.
We will investigate the inverse statement, that is the fact that no spin-orbit coupling should only be possible when the YT chiral angle is zero.

\subsection{Zero YT angle and spin-orbit decoupling}
A zero YT angle implies, from (\ref{beta}), that $r^\mu \mathfrak{s}_\mu=0$. With this constraint in mind, let us choose the matrix spinor in the form
\begin{align}
\Psi&=\sqrt{\rho}\mathcal{B}Ue^{-\gamma_2\gamma_1\frac{\epsilon t}{\hbar}},\label{GenSpinorbetazero}\\
\mathcal{B}&=e^{\left(-\sin\phi\gamma_1\gamma_0+\cos{\phi}\gamma_2\gamma_0\right)\frac{\arctanh{f}}{2}},\\
U&=e^{-\gamma_2\gamma_1\phi/2}e^{-\gamma_3\gamma_1\theta/2}e^{-\gamma_3\gamma_1\pi/4}\label{GenUbetazero}.
\end{align}
In comparing the matrix spinor (\ref{GenSpinor}) against (\ref{GenSpinorbetazero}) we note that, apart from $\sqrt{\rho}$ and the fact that $\beta$ is gone, the only change is in the rotation matrix $U$.

As an illustration, let us choose $f=-\sqrt{1-a^2}$, with $a>0$ and $a<1$ constant. Moreover, let us make the following choice of density
\begin{align}\label{densityzeroYT}
\sqrt{\rho}=\frac{\kappa\left(1-a^2\right)^{1/4}e^{-G/2}}{\sqrt{2a}\,r\sqrt{\sin\theta}}
\end{align}
where $\kappa$ is a normalization constant. Given these choices, from (\ref{GenSpinorbetazero}) we extract the Dirac spinor
\begin{align}\label{spinorCou}
\psi=\sqrt{\rho}e^{-\frac{i\epsilon t}{\hbar}}\left(
\begin{array}{c}
 \frac{(1+a)^{1/2} e^{-\frac{i \phi }{2}} \left(\cos{\frac{\theta}{2}}
 -\sin{\frac{\theta}{2}}\right)}{2}
 \\
 \frac{(1+a)^{1/2} e^{\frac{i \phi }{2}} \left(\sin{\frac{\theta}{2}}
 +\cos{\frac{\theta}{2}}\right)}{2}
 \\
 \frac{i (1-a)^{1/2}e^{-\frac{i \phi }{2}} \left(\sin{\frac{\theta}{2}}
 +\cos{\frac{\theta}{2}}\right)}{2} \\
 -\frac{i (1-a)^{1/2}e^{\frac{i \phi }{2}} \left(\cos{\frac{\theta}{2}}
 -\sin{\frac{\theta}{2}}\right)}{2}
\end{array}
\right)
\end{align}
as easy to see. The most remarkable thing about this spinor is that, from (\ref{scalarpart}) and (\ref{vectorpart}) (RDI is used here because we are still in cartesian coordinates) we extract the following vector potential
\begin{align}
\label{VecPotH}
A^0&=\frac{\sqrt{1-a^2} \hbar G'(r)+\frac{2 a \epsilon }{c}-2 c m}{2 a}\\
A^1&=\frac{\sin\phi \left(-2 \sqrt{1-a^2} c m+\hbar G'(r)\right)}{2 a}\\
A^2&=-\frac{\cos{\phi} \left(-2 \sqrt{1-a^2} c m+\hbar G'(r)\right)}{2 a}\\
A^3&=0
\end{align}
where $r=\sqrt{x^2+y^2+z^2}$, thus implying that in order to remove the spin precession, it was necessary to add a magnetic field. Moreover, with the matrix spinor (\ref{GenSpinorbetazero}) we calculate the following normalized components of the velocity (\ref{velocity1}) and spin (\ref{spin1}) which are given by
\begin{align}
&s_0=0 \ \ \ \ s_1=-\cos{\theta}\cos{\phi} \label{s01EB}\\
&s_2=-\cos{\theta}\sin\phi \ \ \ \ s_3=\sin\theta \label{s23EB}\\
&v_0=\frac{1}{a} \ \ \ \ v_1=-\frac{\sqrt{1-a^2}}{a}\sin\phi \\
&v_2=\frac{\sqrt{1-a^2}}{a}\cos{\phi} \ \ \ \ v_3=0.
\end{align}
Note that the matrix spinor (\ref{GenSpinorbetazero}), the Dirac spinor (\ref{spinorCou}) and the vector potential above satisfy the Dirac equation in \textit{cartesian coordinates}. Moreover, it is noteworthy that the spin vector no longer precess, instead always pointing in the $-\hat{\theta}$ direction. Hence, by making the YT angle equals to zero, the spin-orbit coupling causing the precession of the spin vector also vanishes.

We next change to spherical coordinates.
The form of the spinor (\ref{spinorCou}) remains the same while both $\gamma_a$ and $A_a$ are transformed via the tetrads as usual. Given that all the components of the spin connection $\Omega_{ij\mu}$ with the above tetrads and connection coefficients are zero, the calculation of the components of the $R_{ij\mu}$ tensors is straightforward and the non-zero components are
\begin{align}\label{Rtensor}
&R_{13\theta}=\cos{\phi}\ \ \ \ R_{23\theta}=\sin\phi\\
&R_{12\phi}=-1
\end{align}
We can transform the above tensors to the spacetime manifold as follows
$$
R_{\mu\nu\rho}=e^i_\mu e^j_\nu R_{ij\rho}
$$
whose components are
\begin{align}
&R_{r\theta\theta}=-r\\
&R_{r\phi\phi}=-r|\sin\theta|^2 \ \ \ \ R_{\theta\phi\phi}=-r^2\cos{\theta}\sin\theta
\end{align}
as can be easily checked. Finally, the components of the velocity and spin vectors in spherical coordinates are
\begin{align}
&s_\theta=-r \\
&v_t=\frac{1}{a} \ \ \ \ v_\phi=\frac{\sqrt{1-a^2}}{a}r\sin\theta\label{velocity}
\end{align}
with vector potential
\begin{align}
&A_t=\hbar\frac{\sqrt{1-a^2} G'(r)}{2 a}-\frac{c m}{a }+\frac{\epsilon }{c }\label{A0} \\
& A_\phi=\frac{r \sin{\theta} \left(-2 \sqrt{1-a^2} c m+\hbar G'(r)\right)}{2 a }\label{Aphi}.
\end{align}
and as it is easy to see, the density (\ref{densityzeroYT}) is a solution to the Dirac equation in polar form.

Let us next check some particular solutions.
We first consider the limit $a\rightarrow1$. In doing so (\ref{spinorCou}) and (\ref{velocity}-\ref{Aphi}) become
\begin{align}\label{spinorCoua1}
\psi=\sqrt{\rho}e^{-\frac{imc^2 t}{\hbar}}\left(
\begin{array}{c}
 \frac{ e^{-\frac{i \phi }{2}} \left(\cos{\frac{\theta}{2}}
 -\sin{\frac{\theta}{2}}\right)}{\sqrt{2}}
 \\
 \frac{ e^{\frac{i \phi }{2}} \left(\sin{\frac{\theta}{2}}
 +\cos{\frac{\theta}{2}}\right)}{\sqrt{2}}
 \\
 0 \\
0
\end{array}
\right)
\end{align}
and
\begin{align}
&v_t=1 \ \ \ \ v_\phi=0\\
&A_t=0 \ \ \ \ A_\phi=\frac{r \sin{\theta}\hbar G'(r)}{2}\label{vectpota1}
\end{align}
where we also assumed $\epsilon=mc^2$ for simplicity. In the same limit the density (\ref{densityzeroYT}) would go to zero, but the density is defined only up to a normalization constant. In our case we will choose the following density so to avoid the degeneracy of this limit
\begin{align}\label{densitya1}
\sqrt{\rho}=\frac{\kappa e^{-G/2}}{r\sqrt{2\sin\theta}}
\end{align}
and thus preserving the fact that we still have a solution of the Dirac equations. This special solution for the magnetic-solenoid field is quite interesting. For a discussion on solutions to the Dirac equation in similar fields see \cite{Bagrov2012} and references therein.

As another particular solution which is presented here for the first time, let us consider again (\ref{spinorCou}) but now taking the function $G$ to be
\begin{align}
G=-\frac{2 a \alpha }{\sqrt{1-a^2}} \log \left(\frac{c m r}{\hbar }\right)+\frac{2
 \sqrt{1-a^2} c m r}{\hbar }+\frac{a \mathfrak{i} \mu_0 r}{2 R \hbar }.
\end{align}
This particular form of the function $G$ leads to the following components of the vector potential 
\begin{align}
A_t=-\frac{\alpha \hbar }{r},\,\, A_\phi=\frac{1}{4} \sin{\theta}\left(\frac{\mathfrak{i}\mu_0 r}{R}-\frac{4 \alpha \hbar
 }{\sqrt{1-a^2}}\right).
\end{align}
Thus, such a choice leads to the analytical solution for the ground state of the Hydrogen atom along with a magnetic field having components
\begin{align}\label{MagF}
&B_r=\frac{\mathfrak{i} \mu_0 \cos{\theta}}{2 R}-\frac{2\alpha\hbar\cos{\theta}}{\sqrt{1-a^2} r},\nonumber\\
&B_\theta=\frac{\alpha\hbar\sin{\theta}}{\sqrt{1-a^2} r}-\frac{\mathfrak{i}\mu_0\sin{\theta}}{2 R}
\end{align}
corresponding to a superposition of a magnetic-solenoid field such as the one at the centre of a circular loop of radius $R$ carrying a current $\mathfrak{i}$ and a magnetic field generated by the current density $\mathbf{J}$ 
having only the following component in the azimuthal direction
\begin{align}\label{Current}
J_\phi=-\frac{2 \alpha\hbar \sin{\theta}}{\sqrt{1-a^2} r^2 }.
\end{align}
Incidentally, by making the following substitution 
$$
G=\frac{\mathfrak{i} \mu_0}{2 R \hbar }r
$$ 
into the Dirac spinor (\ref{spinorCoua1}) and the vector potential (\ref{vectpota1}) we arrive at the analytical solution for the case with only the constant and homogeneous magnetic field, i.e., the field one get by putting $\alpha=0$ in (\ref{MagF}). It is noteworthy that this solution corresponds to the three dimensional generalization of the inhomogeneous magnetic field solution given in \cite{RDI3} as can be easily seem if one writes the correspondingly magnetic field in cartesian coordinates and choose $\theta=\pi/2$ (i.e., $z=0$).
\section{Discussion}\label{discussion}
In a comparison between the matrix spinors (\ref{GenSpinor2}) and (\ref{GenSpinorbetazero}), the most noteworthy feature of these solutions to the Dirac equation are the effects the removal of the YT angle have on the physics of the problem. The first notable effect is the change from a purely spherically symmetric electric field in (\ref{GenSpinor2}) to a combination of a spherically symmetric electric field with magnetic fields in (\ref{GenSpinorbetazero}). The second, even more remarkable, effect is that the removal of the YT angle leads to the absence of spin-orbit coupling. That the latter is the case can be clearly inferred from the structure of the matrix spinor as discussed below. 

Regarding the first effect one can note the following. For all localized solutions found using RDI, a matrix spinor with zero YT angle always led to a solution of the Dirac equation for an electron interacting with a magnetic field. It is well-known that for the case of a magneto-static field there exists a Foldy-Wouthuysen transformation in closed form that exactly diagonalizes the Dirac Hamiltonian \cite{PhysRev.111.1011}, thus allowing a full separation into states of positive and negative energy (or charge). This is consistent with the interpretation that a non-zero YT angle corresponds to a mixture of positive and negative energy states, even more so in the Hydrogen-like atom case in which the YT angle depends on $\theta$. The form of the matrix spinor (\ref{GenSpinor2}) and the observation that for stationary electric fields no closed form Foldy-Wouthuysen transformation exists (yet) which exactly diagonalizes the Dirac Hamiltonian led us to speculate that, in such cases, a non-zero YT angle in the matrix spinor is necessary. However, the matrix spinor (\ref{GenSpinorbetazero}) proves that one can still have an inhomogeneous static electric field for a matrix spinor having zero YT angle with the expense of also adding a magnetic field. These findings suggest a deep connection between positive and negative energy states separability, magnetic fields and the YT angle.

Finally, in the case of the second effect, it seems to be connected with the geometric features of the matrix spinors (\ref{GenSpinor2}) and (\ref{GenSpinorbetazero}). It was previously mentioned that the function $f$ is the magnitude of the electron's velocity. In the case of the matrix spinor (\ref{GenSpinor2}) for constant $X$ (i.e., the spinor corresponding to the ground state of a Hydrogen-like atom) the spatial components of the velocity four-vector $v\!\!\!/$ live on the circles of latitude of a sphere of radius $Z\alpha$ as can be inferred from their dependence on $\sin\theta$. In contrast, for the matrix spinor (\ref{GenSpinorbetazero}) the electron's velocity is everywhere constant and has its direction opposite to the Hydrogen-like atom case. Moreover, from the definition of the spin vector (\ref{spin1}) we see that the considerable change of the rotation matrix $U$ from (\ref{GenSpinor2}) to (\ref{GenSpinorbetazero}) will greatly influence the form of $s\!\!\!/$. This implies that one effect of the YT angle is to tilt the spin vector as can be seen by the appearance of the term $\tan(\beta/2)$ in the rotation matrix (\ref{GenU}), 
making it dependent on the boost matrix in the Hydrogen-like case in contrast to the electric plus magnetic field case where the rotation matrix (\ref{GenUbetazero}) is such that $U\gamma_3U^\dagger$ end up commuting with $\mathcal{B}$ in (\ref{spin1}); such a feature can easily be inferred from (\ref{s1E}), (\ref{s2E}), (\ref{s3E}), (\ref{s01EB}) and (\ref{s23EB}). It is this commutativity between the boost and the rotation matrix that led us to conclude the connection between a non-zero YT angle with the spin-orbit coupling. The fact that the aforementioned features are so straightforwardly identifiable in the matrix spinor, while being hidden in the standard Dirac spinor, highlights the advantages of the CRDI technique in the geometrical interpretation of solutions to the Dirac equation.
\section{Conclusion}\label{conclusion}

We have written RDI in explicitly covariant form, thus putting forward the more general CRDI technique, which is the main result of this work. We then showed how it can be used to perform non-trivial change of reference frames with respect to which a given matrix spinor is given that can potentially be a powerful tool in the quest of novel solutions to the Dirac equation. In addition, a whole family of normalizable analytic solutions to the Dirac equation is presented. More specifically, we find exact solutions for the case of a Dirac electron in the presence of a magnetic field as well as a novel solution consisting of a combination of a spherically symmetric electric field with magnetic fields. Giving the connection of the YT angle $\beta$ with quantum states having particle and antiparticle admixtures as well as with the dynamics of the spinor's internal degrees of freedom (a.k.a spin), its role in the solutions to the Dirac equation is yet to be fully elucidated. Hence, an important feature of the solutions presented here is that they offer some hints on a possibly deep connection between $\beta$, magnetic fields and spin-orbit coupling for normalizable Dirac spinors. In fact, we propose the following conjecture: The only \textit{localized} (normalizable) solutions to the Dirac equation having no spin-orbit coupling are those having zero YT angle. Therefore, proving (or disproving) this conjecture would be an important contribution in better understanding the geometrical role of $\beta$ in the Dirac spinor.

Another application of CRDI would be the construction of solutions of the Dirac equation in presence of both electromagnetic and gravitational fields. For instance, such solutions could be used to study the so called traversable wormholes, i.e., stable wormholes that does not require exotic matter \cite{WH1,WH2,WH3}. It is noteworthy the recently proposed solution describing a macroscopic (that is, allowing for humans to travel through it) traversable wormhole \cite{WH4}.
Also, solutions to the Dirac equation in the presence of gravitational fields could be used in order to test the predictions that matter interacting with a quantized gravitational field should lead to an entangling interaction between massive objects \cite{Graviton1}.

\end{document}